\documentclass[preprint,12pt,nofootinbib,groupedaddress,superscriptaddress]{revtex4-1}
\usepackage{graphicx}
\usepackage{dcolumn}
\usepackage{bm}
\usepackage{amssymb}
\usepackage{amsmath}
\usepackage{epsfig}    
\usepackage{color}
\usepackage{slashed}
\usepackage{comment}
\usepackage{hyperref}

\begin{document} 

\preprint{OU-HET-1239}
\preprint{NU-EHET 003}
 
\title{New renormalization scheme in the two Higgs doublet models}

\author{Shinya Kanemura}
\email{kanemu@het.phys.sci.osaka-u.ac.jp}
\affiliation{Department of Physics, Osaka University, Toyonaka, Osaka 560-0043, Japan}

\author{Mariko Kikuchi}
\email{kikuchi.mariko13@nihon-u.ac.jp}
\affiliation{College of Engineering, Nihon University, Koriyama, Fukushima 963-8642, Japan}

\author{Kei Yagyu}
\email{yagyu@het.phys.sci.osaka-u.ac.jp}
\affiliation{Department of Physics, Osaka University, Toyonaka, Osaka 560-0043, Japan}

\begin{abstract}
\noindent
We propose a new renormalization scheme in the two Higgs doublet models with a softly-broken $Z_2$ symmetry and CP-conservation in the Higgs sector. 
In this scheme, counterterms for mixing angles of the Higgs bosons are determined by using the decay rates of the discovered Higgs boson $h$, $i.e.$, $h \to \tau^+\tau^-$
and $h \to ZZ^* \to Z\ell^+\ell^-$ at next leading order (NLO) instead of using the renormalized two-point functions which are adopted in the previous scheme.  
We require that the decay rates at NLO are determined to be the corresponding predictions at NLO in the Standard Model (SM) times square of the scaling factor which describes the deviation of $h$ couplings at tree level from the SM value.  
The mixing angles then maintain the meaning of the ``alignmentness", $i.e.$, 
how the properties of $h$ are close to the SM predictions, while they lose such meaning in the previous scheme.  
We compare the predictions of the decay rates at NLO given in the new scheme and those in the previous scheme. 

 
\end{abstract}

\maketitle
\newpage

\section{Introduction}

Up to now, the measured properties of the discovered Higgs boson $h_{125}$ are consistent with those of the Standard Model (SM) predictions within the theoretical and experimental uncertainties~\cite{ATLAS:2022vkf,CMS:2022dwd}.
This, however, does not necessarily mean that the minimal Higgs sector assumed in the SM is correct. In fact, there is no compelling reason to consider the minimal form, but are possibilities for realizing extended structures of the Higgs sector. 
The latter often appears in various new physics scenarios such as those to explain neutrino masses, dark matter and baryon asymmetry of the Universe.  
Therefore, it is quite important to reconstruct the true structure of the Higgs sector by experiments in order to determine new physics beyond the SM.  

A robust way to extract the structure of the Higgs sector is to  precisely measure the properties of $h_{125}$ such as the coupling constants, the production cross sections and the decay branching ratios. 
The precise measurements of $h_{125}$ can be performed at the High-Luminosity LHC (HL-LHC)~\cite{ATLAS:2013hta,CMS:2013xfa} and future lepton colliders, e.g., the International Linear Collider (ILC)~\cite{Baer:2013cma,Asai:2017pwp,Fujii:2017vwa,ILC:2019gyn}, the Circular Electron-Positron Collider (CEPC)~\cite{CEPC-SPPCStudyGroup:2015csa}, the Future Circular Collider (FCC-ee)~\cite{FCC:2018byv} and the Compact LInear Collider (CLIC)~\cite{Klamka:2021cjt}.
For instance, at the ILC with the collision energy of 250 GeV, the coupling constants of $h_{125}$ are expected to be measured with a percent or less than percent level~\cite{Fujii:2017vwa}. 
Thus, it is inevitable to prepare the calculations beyond the leading order (LO) in order to compare such precise measurements. 
Several numerical tools have been public, which provide precise calculations for the properties of $h_{125}$ in extended Higgs sectors such as {\tt H-COUP}~\cite{Kanemura:2017gbi,Kanemura:2019slf,Aiko:2023xui}, {\tt 2HDECAY}~\cite{Krause:2018wmo}, {\tt Prophecy4f}~\cite{Denner:2019fcr}, {\tt ewN2HDECAY}~\cite{Krause:2019oar}, {\tt EWsHDECAY}~\cite{Egle:2023pbm} and {\tt FlexibleDecay}~\cite{Athron:2021kve}. 

In extended Higgs sectors, the properties of $h_{125}$ are generally modified from the SM predictions through the mixing of Higgs bosons at tree level, so that the mixing angle describes the ``alignmentness" of the Higgs sector. 
Namely, the mixing angle controls how the couplings of $h_{125}$ deviate from the corresponding SM predictions. 
We also can define the so-called Higgs alignment limit by taking an appropriate value of the mixing angle, in which all the deviations of $h_{125}$ couplings vanish at tree level. 

The situation described in the above can drastically be different when we take into account quantum effects on the $h_{125}$ couplings, where various model parameters, e.g., masses of additional Higgs bosons, enter into the calculation via loop diagrams. 
Therefore, the mixing angle generally loses the meaning of the alignmentness at loop levels. 
Basically, this does not cause any problem for the comparison between the precise measurements and the theory predictions including radiative corrections. For instance, we can take sets of model parameters 
which provide a given value of deviations of $h_{125}$. 
This, however, makes it difficult to extract physical implications from the measured deviations. 

In this Letter, we develop a new renormalization scheme such that the mixing angle still works to describe the alignmentness of the Higgs sector at loop levels. 
In order to discuss it concretely, we focus on two Higgs doublet models (2HDMs) with a softly-broken $Z_2$ symmetry, in which two mixing angles, denoted as $\alpha$ and $\beta$, appear in the Higgs boson couplings.  
In our scheme, these mixing angles are renormalized in such a way that the decay rates of $h_{125} \to ZZ^* \to Z\ell^+\ell^-$ and $h_{125} \to \tau^+\tau^-$ at next leading order (NLO) take the corresponding SM predictions at NLO times square of the mixing factor which describes the scaling factor of the $h_{125}\, ZZ$ and $h_{125}\,\tau^+\tau^-$ couplings at tree level. We show that this scheme works in the 2HDMs and demonstrate how the other decay modes are predicted at NLO by using the new scheme.

\section{Model \label{sec:model}}

The Higgs sector of the 2HDMs is composed of the scalar isospin $SU(2)_L$ doublets $\Phi_1$ and $\Phi_2$ with the hypercharge $Y=+1/2$. 
We impose a $Z_2$ symmetry, $\Phi_1 \to \Phi_1,~\Phi_2 \to -\Phi_2$, 
to avoid flavor changing neutral currents at tree level~\cite{Glashow:1976nt}, which is softly-broken via a dimensionful parameter in the Higgs potential introduced just below.  
Throughout this Letter, we assume CP-conservation of the Higgs sector. 

The Higgs potential can then be expressed as 
\begin{align}
V&= m_1^2|\Phi_1|^2 + m_2^2|\Phi_2|^2 - m_3^2(\Phi_1^\dagger \Phi_2 +\text{h.c.}) \notag\\
&+\frac{\lambda_1}{2}|\Phi_1|^4+\frac{\lambda_2}{2}|\Phi_2|^4 + \lambda_3|\Phi_1|^2|\Phi_2|^2 + \lambda_4|\Phi_1^\dagger\Phi_2|^2
+ \frac{\lambda_5}{2}\left[(\Phi_1^\dagger\Phi_2)^2+\text{h.c.}\right], \label{pot_thdm1}
\end{align}
where all the parameters are real and the $m_3^2$ term softly breaks the $Z_2$ symmetry. 
The Higgs fields can be parameterized as
\begin{align}
\Phi_a= \left[\begin{array}{c}
\omega_a^+\\
\frac{1}{\sqrt{2}}(v_a+h_a+iz_a)
\end{array}\right],\hspace{3mm}(a=1,2), \label{Eq:parametrizations}
\end{align}
with $v_a$ being the VEVs related to the Fermi constant $G_F$ via 
$v\equiv \sqrt{v_1^2+v_2^2}=(\sqrt{2}G_F)^{-1/2}\simeq 246$ GeV.
It is convenient to define the Higgs basis to express the mass eigenstates of the scalar fileds:  
\begin{align}
\left(\begin{array}{c}
\Phi_1\\
\Phi_2
\end{array}\right)=
\left(\begin{array}{cc}
\cos\beta & -\sin\beta\\
\sin\beta & \cos\beta
\end{array}\right)
\left(\begin{array}{c}
\Phi\\
\Phi'
\end{array}\right),
\end{align}
where  $\tan\beta = v_2/v_1$, and 
\begin{align}
\Phi=\left[
\begin{array}{c}
G^+\\
\frac{1}{\sqrt{2}}(h_1'+v+ iG^0)
\end{array}\right],\quad
\Phi'=\left[
\begin{array}{c}
H^+\\
\frac{1}{\sqrt{2}}(h_2'+iA)
\end{array}\right]. \label{Higgs-basis}
\end{align}
In this basis, $H^\pm$ and $A$ and $h_a^\prime$ ($a = 1,2$) respectively represent
the physical singly-charged, CP-odd and CP-even Higgs bosons, while 
$G^\pm$ and ($G^0$) are Nambu-Goldstone bosons absorbed into the longitudinal components of the $W^\pm$ ($Z$) bosons.  
Among these physical states, $h_a^\prime$ are generally not the mass eigenstates, and they can be mixed as 
\begin{align}
\begin{pmatrix}
h_1^\prime \\
h_2^\prime
\end{pmatrix} =
\begin{pmatrix}
\cos(\beta-\alpha) & \sin(\beta-\alpha) \\
-\sin(\beta-\alpha) & \cos(\beta-\alpha)
\end{pmatrix}
\begin{pmatrix}
H\\
h
\end{pmatrix}, 
\end{align}
where $h$ can be identified with the discovered Higgs boson. 

The stationary conditions are given by requiring that the tadpoles for $h_a$ 
vanish at tree level:
\begin{align}
T_a \equiv \frac{\partial V}{\partial h_a}\Big|_0 = 0, 
\end{align}
by which we can eliminate the $m_1^2$ and $m_2^2$ parameters. 
The squared masses of the Higgs bosons are then expressed as
\begin{align}
m_{H^\pm}^2=M^2-\frac{v^2}{2}(\lambda_4+\lambda_5),\quad m_A^2&=M^2-v^2\lambda_5, \label{mass1}
\end{align}
where $M^2\equiv m_3^2/(\sin\beta\cos\beta)$.  
%
The squared mass matrix for the CP-even Higgs bosons is given in the Higgs basis ($h_1^\prime,h_2^\prime$) as 
\begin{align}
{\cal M} =v^2 
\begin{pmatrix}
\lambda_1c^4_\beta + \lambda_2 s^4_\beta +\frac{\lambda_{345}}{2}s^2_{2\beta}
& \frac{1}{2}(\lambda_2 s^2_\beta -\lambda_1c^2_\beta + \lambda_{345}c_{2\beta}) s_{2\beta}\\
\frac{1}{2}(\lambda_2 s^2_\beta -\lambda_1c^2_\beta + \lambda_{345}c_{2\beta}) s_{2\beta}& \frac{M^2}{v^2} + \frac{1}{4}(\lambda_1+\lambda_2-2\lambda_{345}) s^2_{2\beta}
\end{pmatrix},  \label{eq:lams}
\end{align}
where $\lambda_{345}\equiv \lambda_3+\lambda_4+\lambda_5$. 
In Eq.~(\ref{eq:lams}) and the following expressions, we use the shorthand notation $c_\theta \equiv \cos\theta$, $s_\theta \equiv \sin\theta$ and $t_\theta \equiv \tan\theta$. 
The mass eigenvalues and the mixing angle are expressed by 
\begin{align}
m_H^2&= {\cal M}_{11}\, c^2_{\beta-\alpha}+{\cal M}_{22}\, s^2_{\beta-\alpha} -{\cal M}_{12}\, s_{2(\beta-\alpha)},\\
m_h^2&= {\cal M}_{11}\,  s^2_{\beta-\alpha}+{\cal M}_{22}\,  c^2_{\beta-\alpha}+{\cal M}_{12}\, s_{2(\beta-\alpha)},\\
\tan 2(\beta-\alpha)&=\frac{2{\cal M}_{12}}{{\cal M}_{22}-{\cal M}_{11}}. \label{Eq:tan2}
\end{align}

The kinetic terms and Yukawa interactions are expressed in the Higgs basis as 
\begin{align}
{\cal L}_{\rm kin} &= |D_\mu \Phi|^2 +  |D_\mu \Phi'|^2, \\
{\cal L}_{\rm Y} &= -
\frac{\sqrt{2}}{v}\left[
\bar{Q}_L M_u (\tilde{\Phi} + \zeta_u \tilde{\Phi}')u_R 
+\bar{Q}_L M_d(\Phi + \zeta_d \Phi')d_R 
+\bar{L}_L M_e(\Phi + \zeta_e \Phi')e_R\right] + \text{h.c.},  
\end{align}
where $D_\mu$ is the covariant derivative, 
$M_f$ ($f=u,d,e$) are the diagonalized mass matrices for fermions, 
$\tilde{\Phi}^{(\prime)} = i\tau_2\Phi^{(\prime)*}$ and $\zeta_f$ are the flavor universal parameters depending on the four types of Yukawa interactions~\cite{Barger:1989fj,Grossman:1994jb,Aoki:2009ha} as follows:
\begin{align}
&\zeta_u = \zeta_d = \zeta_e = \cot\beta~~~~~~~~~~~~~~~~~~(\text{Type-I}), \\
&\zeta_u = \cot\beta,\quad \zeta_d = \zeta_e = -\tan\beta~~~~(\text{Type-II}), \\
&\zeta_u = \zeta_e = \cot\beta,\quad \zeta_e = -\tan\beta~~~~(\text{Type-X}), \\
&\zeta_u = \zeta_e = \cot\beta,\quad \zeta_d = -\tan\beta~~~~(\text{Type-Y}).  
\end{align}
We give the relevant trilinear interactions in the following discussion at tree level as 
\begin{align}
{\cal L}_{\rm int} & \supset g^{\mu\nu}\kappa_V^{\phi}
\left(\frac{2m_W^2}{v}\phi W_\mu^+ W_\nu^- + \frac{m_Z^2}{v}\phi Z_\mu Z_\nu \right) - \frac{m_f}{v}\kappa_f^\phi \phi\bar{f}f \notag\\
& + \lambda_{hhh}hhh + \lambda_{Hhh}Hhh,
\end{align}
where $\phi = h,H$ and 
\begin{align}
\begin{split}
\kappa_V^h &= s_{\beta-\alpha},~~
\kappa_V^H  = c_{\beta-\alpha}, \\
\kappa_f^h &= s_{\beta-\alpha} + \zeta_f c_{\beta-\alpha},~~
\kappa_f^H = c_{\beta-\alpha} - \zeta_f s_{\beta-\alpha}, \\
\lambda_{hhh} & = -\frac{m_h^2}{2v}\left\{ 
s_{\beta-\alpha} + \left[2c_{\beta-\alpha}^2s_{\beta-\alpha} + c_{\beta-\alpha}^3(\cot\beta -\tan\beta)\right]
\left(1 - \frac{M^2}{m_h^2} \right)
\right\}, \\
\lambda_{Hhh}&=-\frac{c_{\beta-\alpha}}{2v}\Big\{4M^2-2m_h^2-m_H^2 \\
& +(2m_h^2+m_H^2-3M^2)[2c^2_{\beta-\alpha}+s_{\beta-\alpha}c_{\beta-\alpha}(\tan\beta-\cot\beta)]\Big\}. \label{eq:tree}
\end{split} 
\end{align}

\section{Renormalization \label{sec:reno}}

\subsection{Renormalization of scalar two-point functions \label{subsec:2p}}

All the bare parameters in the Higgs potential are shifted into the renormalized ones and the counterterms as follows: 
\begin{align}
\begin{split}
&T_a \to 0 + \delta T_a\quad (a = 1,2), \\
& v \to v + \delta v, \\
&m_\varphi^2 \to m_\varphi^2 + \delta m_\varphi^2 \quad (\varphi = h,~H,~A,~H^\pm), \\
& M^2 \to M^2 + \delta M^2, \\
& \alpha \to \alpha + \delta \alpha,\quad 
\beta \to \beta + \delta \beta.
\end{split}
\end{align}
In the following discussion, we adopt the so-called alternative tadpole scheme~\cite{Fleischer:1980ub,Krause:2016oke}, in which 
the tadpole counterterms $\delta T_a$ are set to be zero, while contributions from one-particle irreducible (1PI) diagrams to renormalized vertex functions 
are replaced by the ordinary 1PI contribution and the tadpole inserted diagrams. 
All the bare scalar fields can be shifted as 
\begin{align}
\begin{pmatrix}
H \\
h 
\end{pmatrix}
&\to 
\begin{pmatrix}
1 + \frac{1}{2}\delta Z_H &\delta Z_{Hh} \\
\delta Z_{hH} & 1 + \frac{1}{2} \delta Z_h
\end{pmatrix}\begin{pmatrix}
H \\
h 
\end{pmatrix}, \\
\begin{pmatrix}
G^0 \\
A 
\end{pmatrix}
&\to 
\begin{pmatrix}
1 + \frac{1}{2}\delta Z_{G} &\delta Z_{GA} \\
\delta Z_{AG} & 1 + \frac{1}{2}\delta Z_A
\end{pmatrix}\begin{pmatrix}
G^0 \\
A 
\end{pmatrix}, \\
\begin{pmatrix}
G^\pm \\
H^\pm 
\end{pmatrix}
&\to 
\begin{pmatrix}
1 + \frac{1}{2}\delta Z_{G^\pm} &\delta Z_{G^\pm H^\mp} \\
\delta Z_{H^\pm G^\mp} & 1 + \frac{1}{2}\delta Z_A 
\end{pmatrix}\begin{pmatrix}
G^\pm \\
H^\pm 
\end{pmatrix}, 
\end{align}
where these off-diagonal elements, e.g., $\delta Z_{Hh}$ and $\delta Z_{hH}$ are generally be independent with each other. 
Now, we have 20 counterterms in total, excluding $\delta T_a$.   
It is worth mentioning the relation between these off-diagonal elements and those given in Ref.~\cite{Kanemura:2015mxa} as follows: 
\begin{align}
\begin{split}
\delta Z_{Hh} &= \delta C_{Hh} + \delta \alpha, \quad\quad
\delta Z_{hH} = \delta C_{hH} - \delta \alpha, \\
\delta Z_{GA} &= \delta C_{GA} + \delta \beta, \quad\quad
\delta Z_{AG} = \delta C_{AG} - \delta \beta, \\
\delta Z_{G^\pm H^\mp} &= \delta C_{G^\pm H^\mp} + \delta \beta, \quad
\delta Z_{H^\pm G^\mp} = \delta C_{H^\pm G^\mp} - \delta \beta. 
\end{split} \label{eq:kosy}
\end{align}
In Ref.~\cite{Kanemura:2004mg}, the wavefunction renormalizations 
in the off-diagonal elements are taken to be the same, i.e., $\delta C_{Hh} = \delta C_{hH}$, $\delta C_{GA} = \delta C_{AG}$ and $\delta C_{G^\pm H^\mp} = \delta C_{G^\pm H^\mp}$. 
In addition, the mixing counterterms $\delta\alpha$ and $\delta \beta$ are determined by imposing the renormalization conditions for the scalar two-point functions, while these counterterms are not determined by using the two-point functions in 
our new renormalization scheme explained later. 
The renormalized two-point functions are then expressed as 
\begin{align}
\hat{\Pi}_{\varphi \varphi}(p^2) &= \Pi_{\varphi \varphi}(p^2) + (p^2-m_\varphi^2)\delta Z_\varphi - \delta m_\varphi^2 \quad (\varphi = h,~H,~A,~H^\pm), \\
\hat{\Pi}_{G^0G^0}(p^2) &= \Pi_{G^0G^0}(p^2) + p^2\delta Z_{G^0} ,\\
\hat{\Pi}_{G^+G^-}(p^2) &= \Pi_{G^+G^-}(p^2) + p^2\delta Z_{G^\pm} ,\\
\hat{\Pi}_{Hh}(p^2) &= \Pi_{Hh}(p^2) + (p^2-m_h^2)\delta Z_{hH} + (p^2-m_H^2)\delta Z_{Hh},\\ 
\hat{\Pi}_{G^0A}(p^2) &= \Pi_{GA}(p^2) + (p^2-m_A^2)\delta Z_{AG^0} + p^2\delta Z_{G^0A},\\
\hat{\Pi}_{G^\pm H^\mp}(p^2) &= \Pi_{G^\pm H^\mp}(p^2) + (p^2-m_{H^{\pm}}^2)\delta Z_{H^\pm G^\mp} + p^2\delta Z_{G^\pm H^\mp}, 
\end{align}
where $\Pi_{XY}$ denote the contributions from the 1PI diagrams and the tadpole inserted diagrams. 

We impose the following 16 on-shell conditions: 
\begin{align}
&\hat{\Pi}_{\varphi \varphi}(m_\varphi^2) = 0 , \label{eq:os1}\\ 
&\frac{d\hat{\Pi}_{\varphi \varphi}(p^2)}{dp^2}\Big|_{p^2 = m_\varphi^2} = 
\frac{d\hat{\Pi}_{G^0G^0}(p^2)}{dp^2}\Big|_{p^2 = 0} = 
\frac{d\hat{\Pi}_{G^+G^-}(p^2)}{dp^2}\Big|_{p^2 = 0} = 0 , \label{eq:os2}\\ 
&\hat{\Pi}_{Hh}(m_H^2) = \hat{\Pi}_{Hh}(m_h^2) = \hat{\Pi}_{G^0A}(0) = \hat{\Pi}_{G^0A}(m_A^2) = 
\hat{\Pi}_{G^\pm H^\mp}(0) = \hat{\Pi}_{G^\pm H^\mp}(m_{H^\pm}^2) = 0. \label{eq:os3}
\end{align}
The following 16 counterterms are then determined as 
\begin{align}
\delta m_\varphi^2 &= \Pi_{\varphi \varphi}(m_\varphi^2),\quad
\delta Z_\varphi   = -\frac{d\Pi_{\varphi \varphi}(p^2)}{dp^2}\Big|_{p^2 = m_\varphi^2},\\
\delta Z_{G^0}   &= -\frac{d\Pi_{G^0G^0}(p^2)}{dp^2}\Big|_{p^2 = 0},\quad
\delta Z_{G^\pm}   = -\frac{d\Pi_{G^+G^-}(p^2)}{dp^2}\Big|_{p^2 = 0}, \\
\delta Z_{Hh}   &= \frac{1}{m_H^2 - m_h^2}\Pi_{Hh}(m_h^2), ~~\quad
\delta Z_{hH}    = -\frac{1}{m_H^2 - m_h^2}\Pi_{Hh}(m_H^2), \label{eq:delZbhh}\\
\delta Z_{G^0A}   &= -\frac{1}{m_A^2}\Pi_{G^0A}(m_A^2), \quad\quad\quad
\delta Z_{AG^0}    = \frac{1}{m_A^2}\Pi_{G^0A}(0), \\
\delta Z_{G^\pm H^\mp}   &= -\frac{1}{m_{H^\pm}^2}\Pi_{G^\pm H^\mp}(m_{H^\pm }^2), \quad
\delta Z_{H^\pm G^\mp}    = \frac{1}{m_{H^\pm}^2}\Pi_{G^\pm H^\mp}(0).
\end{align}
We still have the four undertermined counterterms $\delta v$, $\delta M^2$, $\delta \alpha$
and $\delta \beta$. 
The counterterm for the VEV $\delta v$ is determined by imposing the on-shell condition for the electroweak parameters as in the SM~\cite{Bohm:1986rj}: 
\begin{align}
\frac{\delta v}{v} & = 
 \frac{1}{2}\Re
\left[
\frac{s_W^2-c_W^2}{s_W^2}\frac{\Pi_{WW}^T(m_W^2)}{m_W^2}+\frac{c_W^2}{s_W^2}\frac{\Pi^T_{ZZ}(m_Z^2)}{m_Z^2}
- \frac{d}{dp^2}\Pi^T_{\gamma\gamma}(p^2)\Big|_{p^2=0}-\frac{2s_W}{c_W}\frac{\Pi^T_{Z\gamma}(0)}{m_Z^2}\right], \label{del_vev}
\end{align}
where $\Pi_{VV'}^T$ denote the transverse part of the 1PI and tadpole inserted diagrams for the gauge boson two-point functions.  
The counterterm $\delta M^2$ appears in the renormalized scalar trilinear and quartic vertices. One can apply the $\overline{\rm MS}$ scheme to the renormalized $hhh$ vertex 
such that the ultra-violet (UV) divergent part of the $hhh$ vertex is canceled by 
$\delta M^2$~\cite{Kanemura:2004mg}. 
For the remaining two counterterms $\delta\alpha$ and $\delta \beta$, we can use the renormalized Higgs-gauge-gauge vertex and the Yukawa coupling as we will see in the next section. 
We note that in the renormalization scheme developed in Ref.~\cite{Kanemura:2004mg}, 
let us call it as the KOSY scheme, 
the counterterms $\delta \alpha$ and $\delta \beta$ are determined in terms of the scalar two-point functions by using the identification of Eq.~(\ref{eq:kosy}) assuming $\delta C_{ij} = \delta C_{ji}$ as mentioned above. \footnote{Because of three assumptions of $\delta C_{ij} = \delta C_{ji}$, the number of independent counterterms is seventeen, so that we cannot impose one of the renormalization conditions expressed in Eq.~(\ref{eq:os3}). In Ref.~\cite{Kanemura:2004mg}, the condition $\hat{\Pi}_{G^\pm H^\mp}(m_{H^\pm}^2) = 0$ is not imposed.  }
We also note that the determination of $\delta\alpha$ and $\delta\beta$ by using the two-point functions generally introduces the gauge dependence~\cite{Nielsen:1975fs}, so that the pinch-technique has been applied to remove such a gauge dependence~\cite{Krause:2016oke,Kanemura:2017wtm}. 

\subsection{Renormalized vertices}

We can construct the renormalized Higgs three-point functions by using the counterterms introduced in the previous subsection. 
The renormalized $hV^\mu V^\nu$  ($V=W,~Z$) and $h\bar{f}f$ vertices are decomposed into the following form factors:  
\begin{align}
\hat{\Gamma}_{h VV}^{\mu\nu}(p_1^2,p_2^2,q^2) & = g^{\mu\nu}\hat{\Gamma}_{h VV}^1 + \frac{p_1^\nu p_2^\mu}{m_V^2}\hat{\Gamma}_{h VV}^2 + i\epsilon^{\mu\nu\rho\sigma}\frac{p_{1\rho} p_{2\sigma}}{m_V^2}\hat{\Gamma}_{h VV}^3 , \\
\hat{\Gamma}_{hff}(p_1^2,p_2^2,q^2) & = \hat{\Gamma}_{hff}^{\rm S} + \gamma_5\hat{\Gamma}_{hff}^{\rm P}
+ p_1\hspace{-3.5mm}/\hspace{1.5mm}\hat{\Gamma}_{hff}^{{\rm V}_1}
+ p_2\hspace{-3.5mm}/\hspace{1.5mm}\hat{\Gamma}_{hff}^{{\rm V}_2} \notag\\
&+ p_1\hspace{-3.5mm}/\hspace{1.5mm}\gamma_5\hat{\Gamma}_{hff}^{{\rm A}_1}
+ p_2\hspace{-3.5mm}/\hspace{1.5mm}\gamma_5\hat{\Gamma}_{hff}^{{\rm A}_2}
+ p_1\hspace{-3.5mm}/\hspace{1.5mm}p_2\hspace{-3.5mm}/\hspace{1.5mm}\hat{\Gamma}_{hff}^{\rm T}
+ p_1\hspace{-3.5mm}/\hspace{1.5mm}p_2\hspace{-3.5mm}/\hspace{1.5mm}\gamma_5\hat{\Gamma}_{hff}^{\rm PT}, \label{hff-eff}
\end{align}
where $\epsilon^{\mu\nu\rho\sigma}$ is the anti-symmetric tensor. 
The renormalized $hhh$ vertex $\hat{\Gamma}_{hhh}$ is Lorentz-scalar, so that we do not need to decompose it.   
Each form factor and $\hat{\Gamma}_{hhh}$ are further decomposed into the following contributions: 
\begin{align}
\hat{\Gamma}_{hXX}^i(p_1^2,p_2^2,q^2) = \Gamma_{hXX}^{i,{\rm tree}} + \delta\Gamma_{hXX}^i + \Gamma_{hXX}^i(p_1^2,p_2^2,q^2), 
\end{align}
where the index $i$ labels the form factor for the $hVV$ and $hf\bar{f}$ vertices, and the last term represents the contributions from the 1PI and tadpole inserted diagrams. 
The tree level contribution is given by 
\begin{align}
\Gamma_{h VV}^{1,{\rm tree}} &= \frac{2m_V^2}{v}\kappa_V^{h},~~ 
\Gamma_{hff}^{\rm S,tree} = -\frac{m_f}{v}\kappa_f^h,~~
\Gamma_{hhh}^{\rm tree} = 3!\, \lambda_{hhh},  
\end{align}
where the $\kappa$ factors and $\lambda_{hhh}$ are defined in Eq.~(\ref{eq:tree}), and tree-level contributions to all the other form factors are zero.  
The counterterm contribution is given by 
\begin{align}
\delta\Gamma_{h VV}^1 &= \Gamma_{hVV}^{1,\rm tree}\left[ \delta_{\rm SM}^V + \frac{1}{t_{\beta-\alpha}} (\delta(\beta -\alpha) + \delta Z_{Hh})
\right],  \label{eq:del_hvv} \\
\delta \Gamma_{hff}^{\rm S} &= \Gamma_{hff}^{\rm S,tree}\Bigg[\delta_{\rm SM}^f+ \frac{\kappa_f^H}{\kappa_f^h}[\delta (\beta-\alpha) + \delta Z_{Hh}] - \frac{(1 + \zeta_f^2)c_{\beta-\alpha}}{\kappa_f^h }\,\delta \beta \Bigg], \label{eq:del_hff} \\
\delta\Gamma_{hhh}&=6\left(\delta\lambda_{hhh}+\frac{3}{2}\lambda_{hhh}\delta Z_h+\lambda_{Hhh}\delta Z_{Hh}\right), 
\end{align}
where $\delta (\beta-\alpha) \equiv \delta \beta - \delta \alpha$, and the counterterm contribution to all the other form factors are zero. 
In the above expressions, 
$\delta_{\rm SM}^V$ and $\delta_{\rm SM}^f$ are given by: 
\begin{align}
 \delta_{\rm SM}^V  &= \frac{\delta Z_h}{2} - \frac{\delta v}{v} +\frac{\delta m_V^2}{m_V^2} + \delta Z_V, \quad
 \delta_{\rm SM}^f =
\frac{\delta Z_h }{2}-\frac{\delta v}{v}+
\frac{\delta m_f}{m_f} + \delta Z_V^f,  
\end{align}
where each counterterm is determined by using the on-shell conditions as in the SM~\cite{Bohm:1986rj}   
\begin{align}
\delta m_V^2 &= \Re \Pi^T_{VV}(m_V^2), \\
\delta Z_Z   &= \frac{c_W^2-s_W^2}{s_W^2}\Re \left[\frac{\Pi^T_{ZZ}(m_Z^2)}{m_Z^2}-\frac{\Pi^T_{WW}(m_W^2)}{m_W^2}\right] \notag\\
&+\frac{2(c_W^2-s_W^2)}{c_Ws_W}\frac{\Re \Pi^T_{Z\gamma}(0)}{m_Z^2}-\Re \frac{d\Pi^T_{\gamma\gamma}(p^2)}{dp^2}\Big|_{p^2=0}, \\
\delta Z_W  & = \frac{c_W^2}{s_W^2}\Re \left[\frac{\Pi^T_{ZZ}(m_Z^2)}{m_Z^2}-\frac{\Pi^T_{WW}(m_W^2)}{m_W^2}\right]+\frac{2c_W}{s_W} \frac{\Re\Pi^T_{Z\gamma }(0)}{m_Z^2}-\Re \frac{d\Pi^T_{\gamma\gamma}(p^2)}{dp^2}\Big|_{p^2=0}, \\
\frac{\delta m_f}{m_f}&=\Re\Pi_{ff}^V(m_f^2) +\Re\Pi_{ff}^S(m_f^2), \label{del_mf} \\ 
\delta Z_f^V&=-\Re\Pi_{ff}^V(m_f^2) -2m_f^2\Re\left[\frac{d\Pi_{ff}^V(p^2)}{dp^2}\Big|_{p^2=m_f^2}+\frac{d\Pi_{ff}^S(p^2)}{dp^2}\Big|_{p^2=m_f^2}\right]. \label{eq:del_zfv}
\end{align}
In the above expression, 
$\Pi_{ff}^S$ and $\Pi_{ff}^V$ are respectively the scalar and vector part of the fermion two-point functions. 
In $\delta \Gamma_{hhh}$, the counterterm $\delta \lambda_{hhh}$
is obtained by the shift of the tree level coupling $\lambda_{hhh}$ given in Eq.~(\ref{eq:tree}) as 
\begin{align}
&\delta \lambda_{hhh}  = -\lambda_{hhh}\frac{\delta v}{v}
-\frac{s_{\beta-\alpha}}{2}\frac{\delta m_h^2}{v}  
 +c_{\beta-\alpha}^2\left[s_{\beta-\alpha} + \frac{c_{\beta-\alpha}}{2}(\cot\beta -\tan\beta)\right]\left(\frac{\delta M^2}{v}
-\frac{\delta m_h^2}{v}\right)\notag\\
&-\frac{m_h^2}{2v}c_{\beta-\alpha}\left\{
1  +\left[2c_{\beta-\alpha}^2-4s^2_{\beta-\alpha}-3 c_{\beta-\alpha}s_{\beta-\alpha}(\cot\beta -\tan\beta)\right]
\left(1 - \frac{M^2}{m_h^2} \right)\right\}\delta(\beta-\alpha)\notag\\
&+\frac{2m_h^2}{v}\frac{c^3_{\beta-\alpha}}{s_{2\beta}^2}\left(1 - \frac{M^2}{m_h^2} \right)\delta\beta. 
\end{align}

As we see in Eqs.~(\ref{eq:del_hvv}) and (\ref{eq:del_hff}), $\delta \alpha$ and $\delta \beta$ are included in the counterterm for the $hVV$ and $hf\bar{f}$ vertices with the different combination. 
Thus, we can determine these two counterterms by imposing the renormalization conditions on the quantities which involve these vertices. 
Regarding $\delta M^2$, we apply the $\overline{\rm MS}$ scheme as we explained in Sec.~\ref{subsec:2p}. 
As we will see in the next section, the UV divergent part of the counterterms $\delta \alpha$ and $\delta \beta$ in the new scheme is the same as that in the KOSY scheme, so that we obtain the same expression for $\delta M^2$ determined by the $\overline{\rm MS}$ scheme.

\section{Decay rates at NLO}

In this section, we consider the decay rates of $h \to f\bar{f}$ and 
$h \to VV^* \to Vf\bar{f}$ ($f \neq t$, $V = W,Z$) at NLO. 
These decay rates are expressed as 
\begin{align}
\Gamma_{\rm NLO}(h\to f\bar{f})&=\Gamma_0(h\to f\bar{f}) (1 + \Delta_{\text{EW}}^f +  \Delta_{\text{QCD}}^f ), \label{eq:hff_nlo}\\
\Gamma_{\rm NLO}(h \to Vf\bar{f}) &= \Gamma_{0}(h \to Vf\bar{f})
(1 + \Delta_{\rm EW}^{Vff} + \Delta_{\rm QCD}^{Vff}), \label{eq:hvv_nlo}
\end{align} 
where $\Delta_{\text{EW}}^X$ and $\Delta_{\text{QCD}}^X$ ($X=f,Z,W$) denote the EW and QCD corrections, respectively. 
In Eqs.~(\ref{eq:hff_nlo}) and (\ref{eq:hvv_nlo}), $\Gamma_0$ denotes the decay rate at LO, see e.g., Refs.~\cite{Kanemura:2019kjg} for the analytic expressions.    
%
The EW corrections are given by 
\begin{align}
\Delta_{\rm EW}^f
& =
2 \frac{\delta \Gamma_{hff}^{\rm S} +  \Gamma_{hff}^{\rm S,\rm 1PI}\Big|_{\rm div} }{\Gamma_{hff}^{\rm S,\rm tree}} + \Delta_{\rm rem}^{f}, \\ 
\Delta_{\rm EW}^{Vff}
& =
2 \frac{\delta \Gamma_{hVV}^1 +  \Gamma_{hVV}^{1,\rm 1PI}\Big|_{\rm div} }{\Gamma_{hVV}^{1,\rm tree}} + \Delta_{\rm rem}^{Vff}, 
\end{align}
where $X|_{\rm div}$ denotes the UV divergent part of the quantity $X$ which is canceled by the corresponding counterterm.
In the above expression, $\Delta_{\rm rem}$ are the UV finite remaining part of the EW correction. 
See Ref.~\cite{Kanemura:2019kjg} for the analytic expressions for 
$\Delta_{\rm rem}^f$ and $\Delta_{\rm rem}^{Vff}$, where the latter involves not only the $hVV$ vertex corrections but also the other corrections such as the contributions from box diagrams.

Now, let us discuss the determination of the counterterms $\delta(\beta - \alpha)$ and $\delta \beta$. 
We impose the following renormalization conditions 
\begin{align}
\Delta_{\rm EW}^{Z\ell\ell} = \Delta_{\rm EW}^{Z\ell\ell}\big|_{\rm SM}, \quad
\Delta_{\rm EW}^\tau = \Delta_{\rm EW}^\tau\big|_{\rm SM}. \label{eq:new-conditions}
\end{align}
Substituting these expressions into Eqs.~(\ref{eq:hff_nlo}) and 
(\ref{eq:hvv_nlo}), we obtain  
\begin{align}
\Gamma_{\rm NLO}(h \to V\ell\bar{\ell}) &=  (\kappa_V^h)^2\times\Gamma_{\rm NLO}(h \to V\ell\bar{\ell})_{\rm SM}, \\
\Gamma_{\rm NLO}(h\to \tau^+\tau^-)&=(\kappa_\tau^h)^2\times
\Gamma_{\rm NLO}(h\to \tau^+\tau^-)_{\rm SM}, 
\end{align}  
where the QCD corrections $\Delta_{\rm QCD}^{\tau}$ and 
$\Delta_{\rm QCD}^{Z\ell\ell}$
are taken to be zero for the above processes. 
This means that the parameters $\kappa_V^h$ and $\kappa_\tau^h$ measure the deviation in the decay rates of $h \to Z\ell\ell$ and $h \to \tau^+\tau^-$ processes, respectively at one-loop level. 
These conditions fix the counterterm as 
\begin{align}
\delta(\beta-\alpha) &= 
- \delta Z_{Hh} 
-t_{\beta-\alpha}\left[
\delta_{\rm SM}^Z
+\frac{\Gamma_{hZZ}^{1,\rm 1PI}\Big|_{\rm div}}{\Gamma_{hZZ}^{1, \rm tree}}
+\frac{\Delta_{\rm rem}^{Z\ell\ell} -\Delta_{\rm EW}^{Z\ell\ell}\big|_{\rm SM}}{2}
\right], \label{eq:deltheta1}\\
\delta \beta &= \frac{\kappa_\tau^h}{(1+\zeta_\tau^2)c_{\beta-\alpha}}\left[
\delta_{\rm SM}^\tau + \frac{\Gamma_{h\tau\tau}^{\rm S,1PI}\big|_{\rm div}}{\Gamma_{h\tau\tau}^{\rm S,tree}} + \frac{\Delta_{\rm rem}^{\tau} - \Delta_{\rm EW}^{\tau}\big|_{\rm SM}}{2}
+ \frac{\kappa_\tau^H}{\kappa_\tau^h}[\delta(\beta-\alpha) + \delta Z_{Hh}]\right].  
\label{eq:deltheta}
\end{align}
Since we use the physical quantities for the renormalization conditions, the gauge dependence does not enter into the above counterterms for the mixing parameters. 
We can check that the UV divergent part of the above counterterms is the same as that given by the KOSY scheme as follows: 
\begin{align}
\delta\alpha|_{\rm div} 
& = \delta\alpha|_{\rm div}^{\rm KOSY}, 
\quad 
\delta \beta|_{\rm div}
= \delta \beta|_{\rm div}^{\rm KOSY}. 
\end{align}

Using the counterterms $\delta(\beta-\alpha)$ and $\delta \beta$ given in Eqs.~(\ref{eq:deltheta1}) and (\ref{eq:deltheta}), the NLO corrections are determined as follows. 
For the $h\to WW^*$ process,  
\begin{align}
\Delta_{\rm EW}^{Wff} & = 
2\left(\frac{\delta m_W^2}{m_W^2} - \frac{\delta m_Z^2}{m_Z^2}
+ \delta Z_W -  \delta Z_Z 
\right)_{\rm fin} + \Delta_{\rm rem}^{Wff} - \Delta_{\rm rem}^{Z\ell\ell} + \Delta_{\rm EW}^{Z\ell\ell}\big|_{\rm SM}, 
\end{align}
where $X|_{\rm fin}$ deonotes the UV finite part of the quantity $X$. 
We see that the contribution from $\delta Z_h$ vanishes in the above expression due to the renormalization condition~(\ref{eq:new-conditions}), which can give a dominant contribution to the $hWW$ vertex, and it is proportional to 
the quadratic power-like dependence of the additional Higgs boson mass for $M^2/v^2 \ll 1$~\cite{Kanemura:2015mxa}. 
Thus, the amount of the deviation in the decay rate of $h \to WW^*$ from the SM prediction is typically smaller than that in the KOSY scheme. 
For the $h \to f\bar{f}$ decays, the expression for $\Delta_{\rm EW}^f$ takes quite a simple form if $\zeta_f = \zeta_\tau$ (the Type-I 2HDM satisfies this condition for any fermion type $f$) as follows:
\begin{align}
\Delta_{\rm EW}^{f} & = 
2\left(\frac{\delta m_f}{m_f} - \frac{\delta m_\tau}{m_\tau}
+\delta Z_V^f - \delta Z_V^\tau  
\right)_{\rm fin} 
+ \Delta_{\rm rem}^{f} - \Delta_{\rm rem}^{\tau} + \Delta_{\rm EW}^{\tau}|_{\rm SM}. \label{eq:same} 
\end{align}
Similar to the case of $\Delta_{\rm EW}^{Wff}$, the contribution from $\delta Z_h$ vanishes in the above formula.  
For the case with $\zeta_f = -1/\zeta_\tau$, we obtain 
\begin{align}
\Delta_{\rm EW}^f &= 
2 \Bigg\{
\frac{\delta m_f}{m_f} -\bar{\zeta}_f \frac{\kappa_\tau^h}{\kappa_f^h}\frac{\delta m_\tau}{m_\tau} + \delta Z_V^f -\bar{\zeta}_f \frac{\kappa_\tau^h}{\kappa_f^h}\delta Z_V^\tau\notag\\
&+\left[1 -\bar{\zeta}_f \frac{\kappa_\tau^h}{\kappa_f^h}
- t_{\beta-\alpha} \left(\frac{\kappa_f^H}{\kappa_f^h} -\bar{\zeta}_f\frac{\kappa_\tau^H}{\kappa_f^h}\right) \right]\left(\frac{\delta Z_h}{2} - \frac{\delta v}{v} \right) \notag\\
&-t_{\beta-\alpha}\left(\frac{\kappa_f^H}{\kappa_f^h} -\bar{\zeta}_f\frac{\kappa_\tau^H}{\kappa_f^h}\right)
\left(\frac{\delta m_Z^2}{m_Z^2}+ \delta Z_Z  \right)\Bigg\}_{\rm fin} \notag\\
& +\Delta_{\rm rem}^f - \bar{\zeta}_f \frac{\kappa_\tau^h}{\kappa_f^h}(\Delta_{\rm rem}^\tau
- \Delta_{\rm EW}^\tau |_{\rm SM}
)
-t_{\beta-\alpha}\left(\frac{\kappa_f^H}{\kappa_f^h} -\bar{\zeta}_f\frac{\kappa_\tau^H}{\kappa_f^h}\right) (\Delta_{\rm rem}^{Z\ell\ell} -\Delta_{\rm EW}^{Z\ell\ell}|_{\rm SM} ), \label{eq:diff}
\end{align}
where 
\begin{align}
\bar{\zeta}_f & \equiv \frac{1 + \zeta_f^2}{1 + \zeta_\tau^2}. 
\end{align}
We can check that Eq.~(\ref{eq:diff}) coincides with Eq.~(\ref{eq:same}) by taking $\zeta_f = \zeta_\tau$ which leads to $\kappa_f^\phi = \kappa_\tau^\phi$ ($\phi = h,H$) and $\bar{\zeta}_f = 1$. 
Differently from the case with $\zeta_f = \zeta_\tau$, the dependences of $t_{\beta-\alpha}$ and $\delta Z_h$ appear, so that $\Delta_{\rm EW}^f$ can be large when we take $s_{\beta-\alpha}\simeq 1$ and/or $M^2 \ll v^2$. 

Now, we show the numerical results of the decay rates of the Higgs boson $h$ at NLO. 
In order to express the deviation from the SM prediction, we introduce
\begin{align}
\Delta R(h \to XY) \equiv  \frac{\Gamma(h \to XY)_{\rm NLO}}{\Gamma(h \to XY)_{\rm NLO}^{\rm SM}} - 1. \label{eq:delr}
\end{align}
For the $h \to ZZ^*$ and $h \to WW^*$ decays,  
we sum all the possible combinations of fermions coming from the decay of the virtual gauge boson. 
We employ the {\tt H-COUP} packages~\cite{Kanemura:2017gbi,Kanemura:2019slf,Aiko:2023xui} for the numerical evaluation of the decay rate at NLO with modifications of the counterterms of $\delta \alpha$ and $\delta \beta$. We apply the highest QCD corrections and the input values of the SM parameters, which are implemented in {\tt H-COUP} Ver. 3~\cite{Aiko:2023xui}.

\begin{figure}[!t]
\begin{center}
\includegraphics[width=72mm]{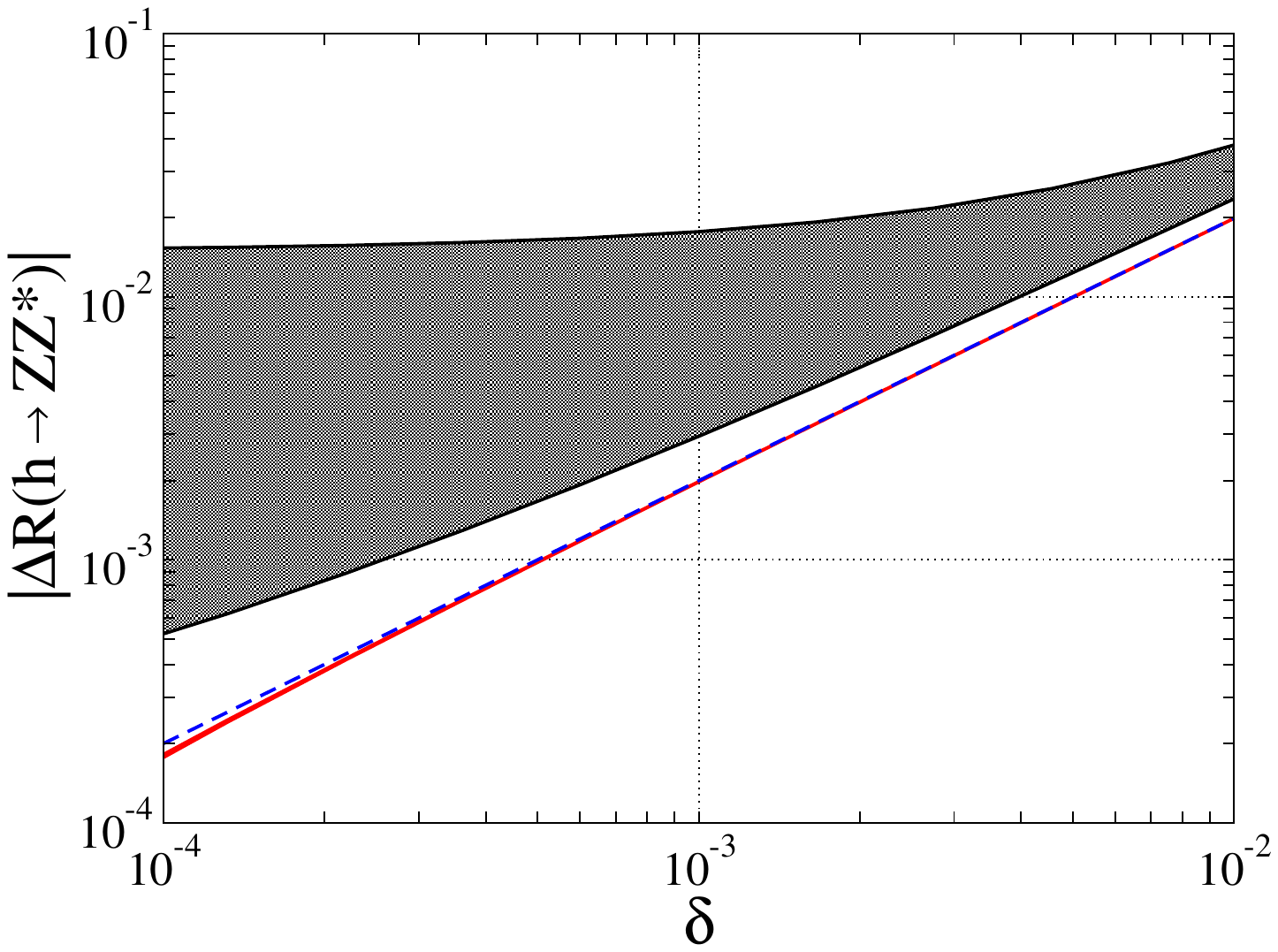}
\includegraphics[width=72mm]{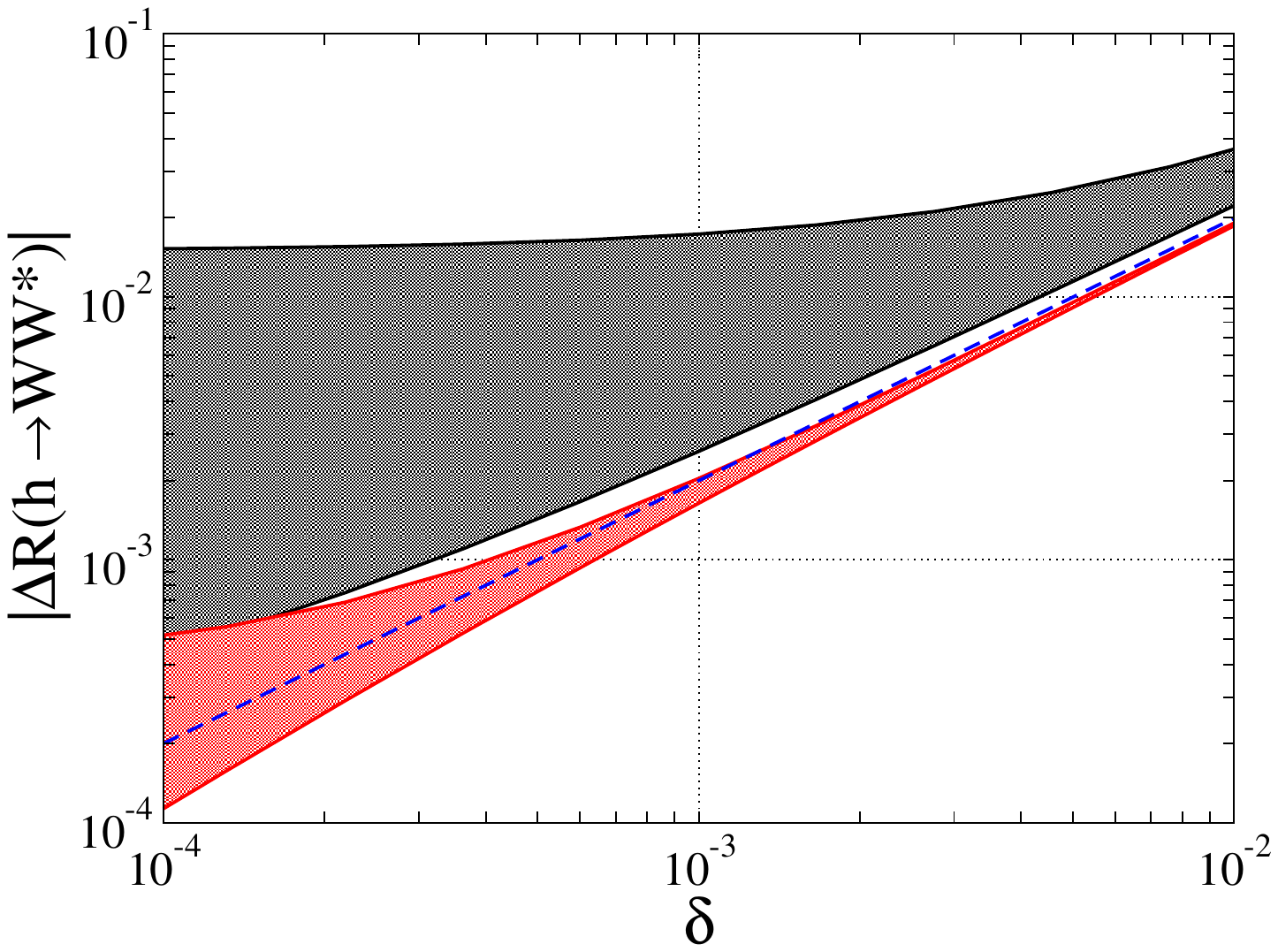}\\
\vspace{-0.1cm}
\includegraphics[width=72mm]{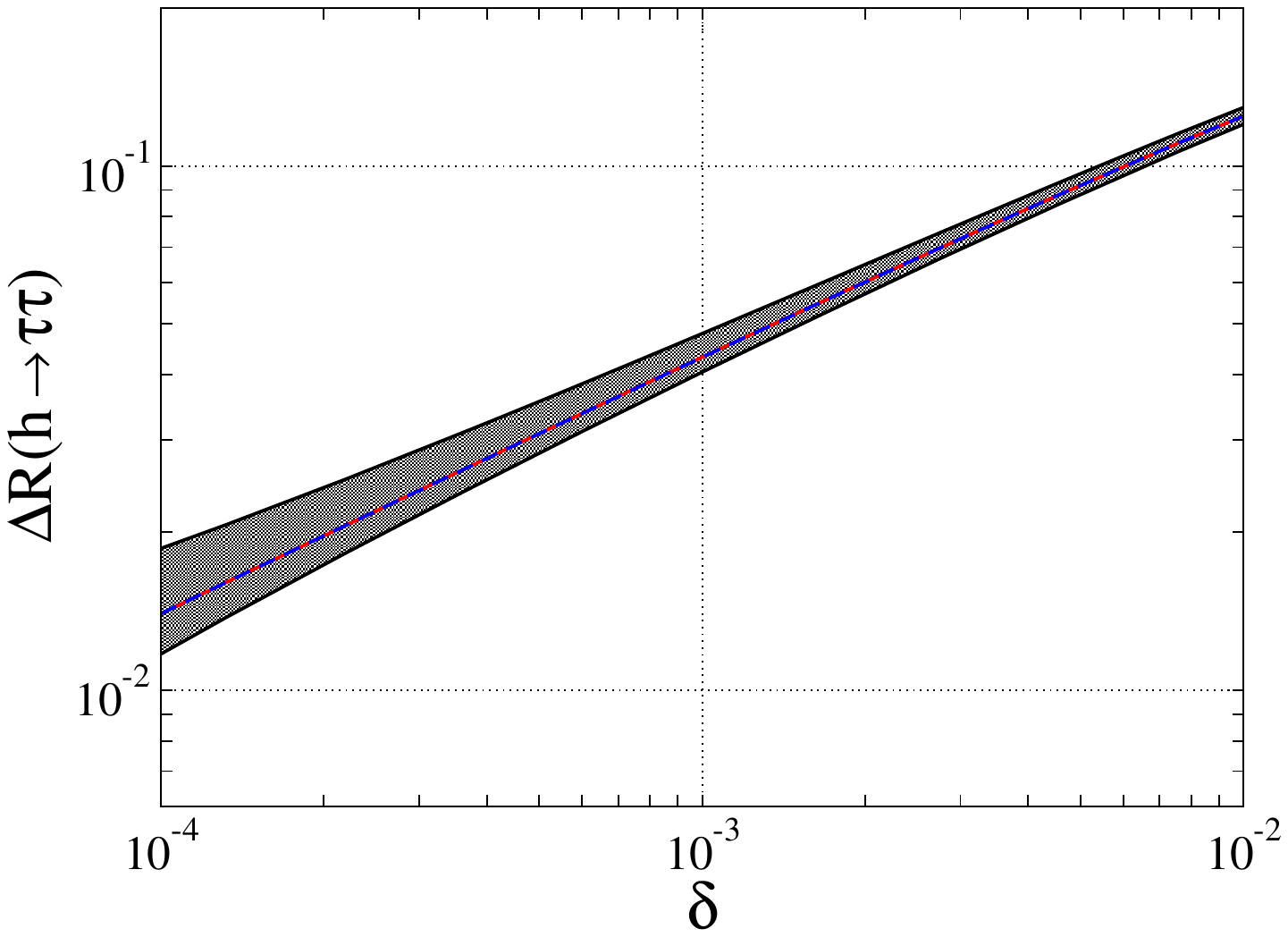}
\includegraphics[width=72mm]{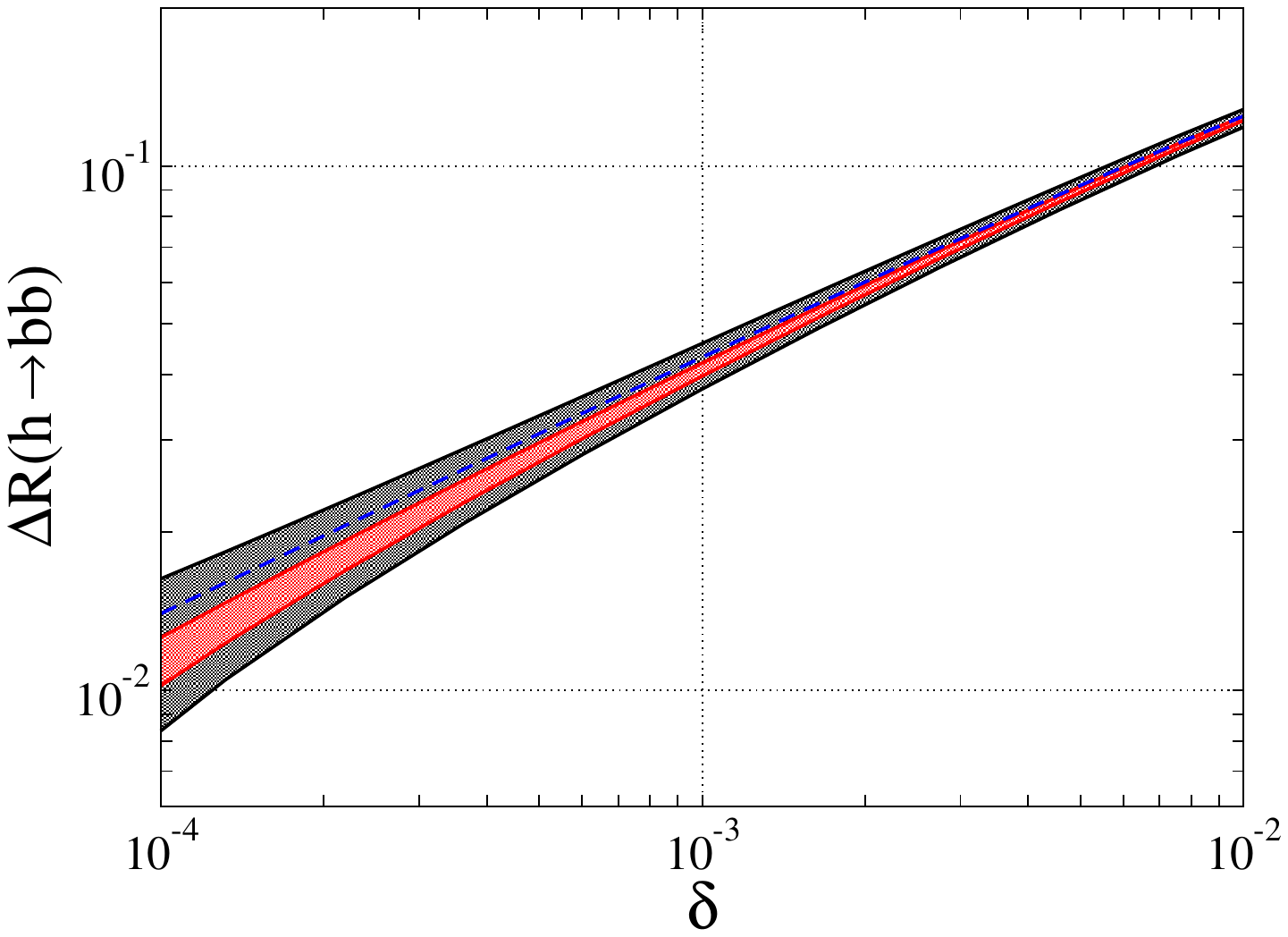}\\
\vspace{-0.1cm}
\includegraphics[width=72mm]{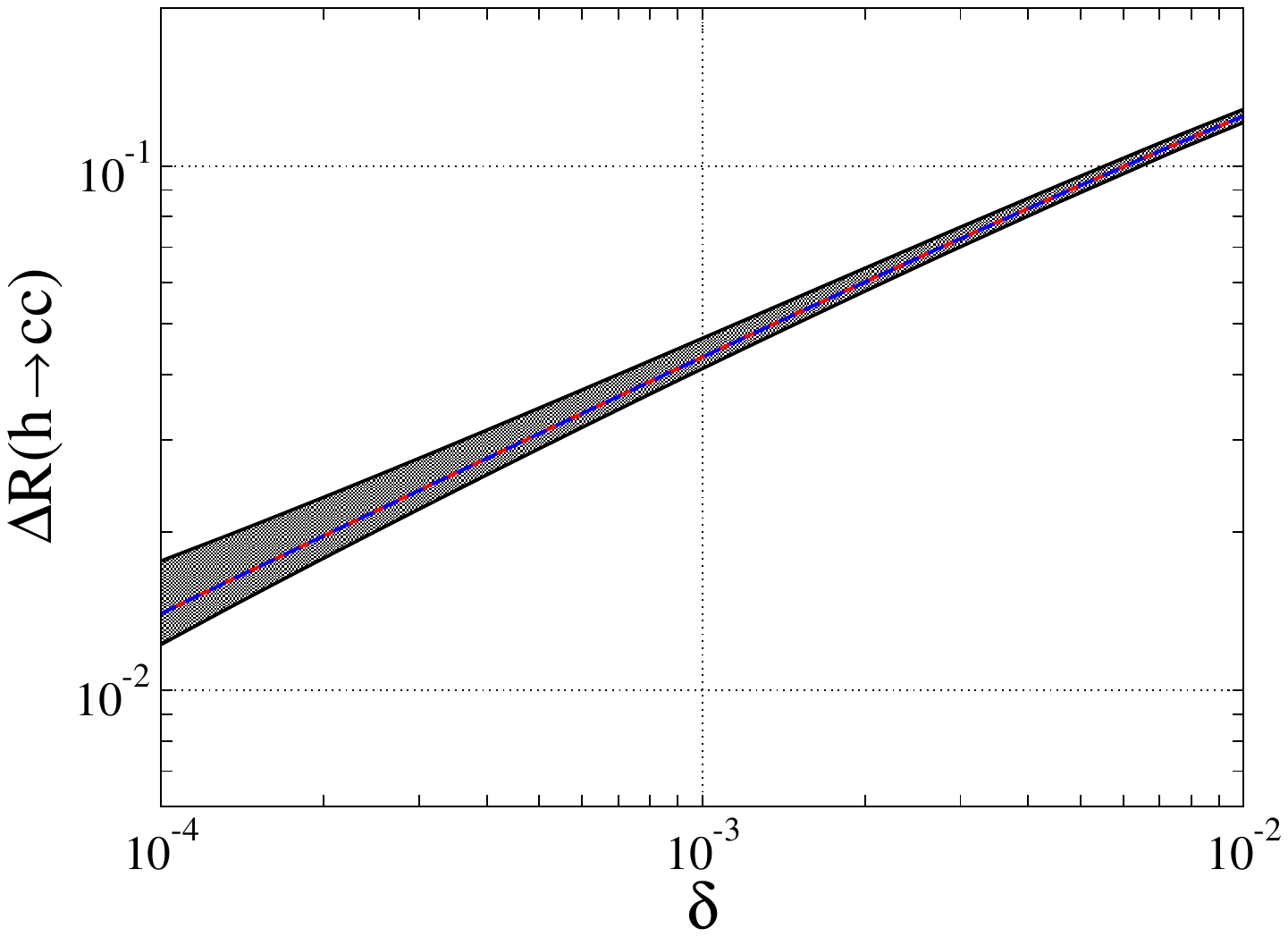}
\end{center}
\vspace{-0.7cm}
\caption{Results of $\Delta R(h \to XY)$, defined in Eq.~(\ref{eq:delr}), with $XY = ZZ^*$ (top-left), $WW^*$ (top-right), $\tau^+\tau^-$ (middle-left), $b\bar{b}$ (middle-right) and $c\bar{c}$ (bottom) as a function of $\delta (\equiv 1 - \sin(\beta-\alpha))$ in the Type-I 2HDM with $m_{H^\pm} = m_H = m_A = 300$ GeV, $\tan\beta = 2$, $\cos(\beta-\alpha) > 0$ and $M^2$ being scanned.
The regions shaded with black and red respectively show the prediction by using the KOSY scheme and the new scheme. The blue dashed curve shows the prediction at LO. For $\Delta R(h \to ZZ^*)$ and $\Delta R(h \to WW^*)$, the results take negative values, so their absolute values are shown. 
}
\label{fig1}
\end{figure}

In Fig.~\ref{fig1}, we show the values of $\Delta R(h \to XY)$ as a function of $\delta \equiv 1 - s_{\beta-\alpha}$ in the Type-I 2HDM with the input parameters written in the caption. We here compare the results at NLO given in the KOSY scheme (black shaded region), those given in the new scheme (red shaded region), and the results at LO (blue dashed curve), 
where the regions are generated due to the scan of the $M^2$ parameter with its region 
being determined by the constraints from the perturbative unitarity~\cite{Kanemura:1993hm,Akeroyd:2000wc,Ginzburg:2005dt,Kanemura:2015ska} and the vacuum stability~\cite{Deshpande:1977rw,Nie:1998yn,Kanemura:1999xf}~\footnote{We take $M^2$ to be positive because the case with $M^2 < 0$ is disfavored by the condition to avoid wrong vacua~\cite{Branchina:2018qlf}.}.
For all the plots, the results in the KOSY scheme give the larger region of the prediction compared with those in the new scheme, because of the contribution from $\delta Z_h$. Typically, the results in the KOSY scheme differ from the LO results by a few percent level.  
On the other hand, the results given in the new scheme show quite good agreement with the LO results, especially for  
$\Delta R(h \to ZZ^*)$ and $\Delta R(h \to \tau^+\tau^- )$ 
due to the renormalization condition.
We also see that $\Delta R(h \to c\bar{c})$ given in the new scheme and the LO shows good agreement with each other. 
We can see the difference between 
the red and blue curves in $\Delta R(h \to WW^*)$, especially for the small $\delta$ region, but the amount of the difference 
is of order $10^{-3}$-$10^{-4}$. 
Therefore, $\kappa_V^h$ works to describe the deviation in the $h \to VV^*$ decays at one-loop level.  
For the $h \to b\bar{b}$ decay, we see a relatively larger difference (but still a few per-mill level) between the results in the new scheme and those at LO. 
This is because the top-quark and $H^\pm$ loop contribution to the $h b\bar{b}$ vertex is not absorbed by the renormalization condition.  
We note that the result of $\Delta R(h \to ZZ^*)$ using the new scheme is not exactly the same as that given at LO, because the hadronic final states ($h \to Zq\bar{q},~q\neq t$) are not used as the renormalization condition. In fact, we see a tiny difference between the results in the new scheme and those at LO at $\delta \sim 10^{-4}$.

\begin{figure}[!t]
\begin{center}
\includegraphics[width=75mm]{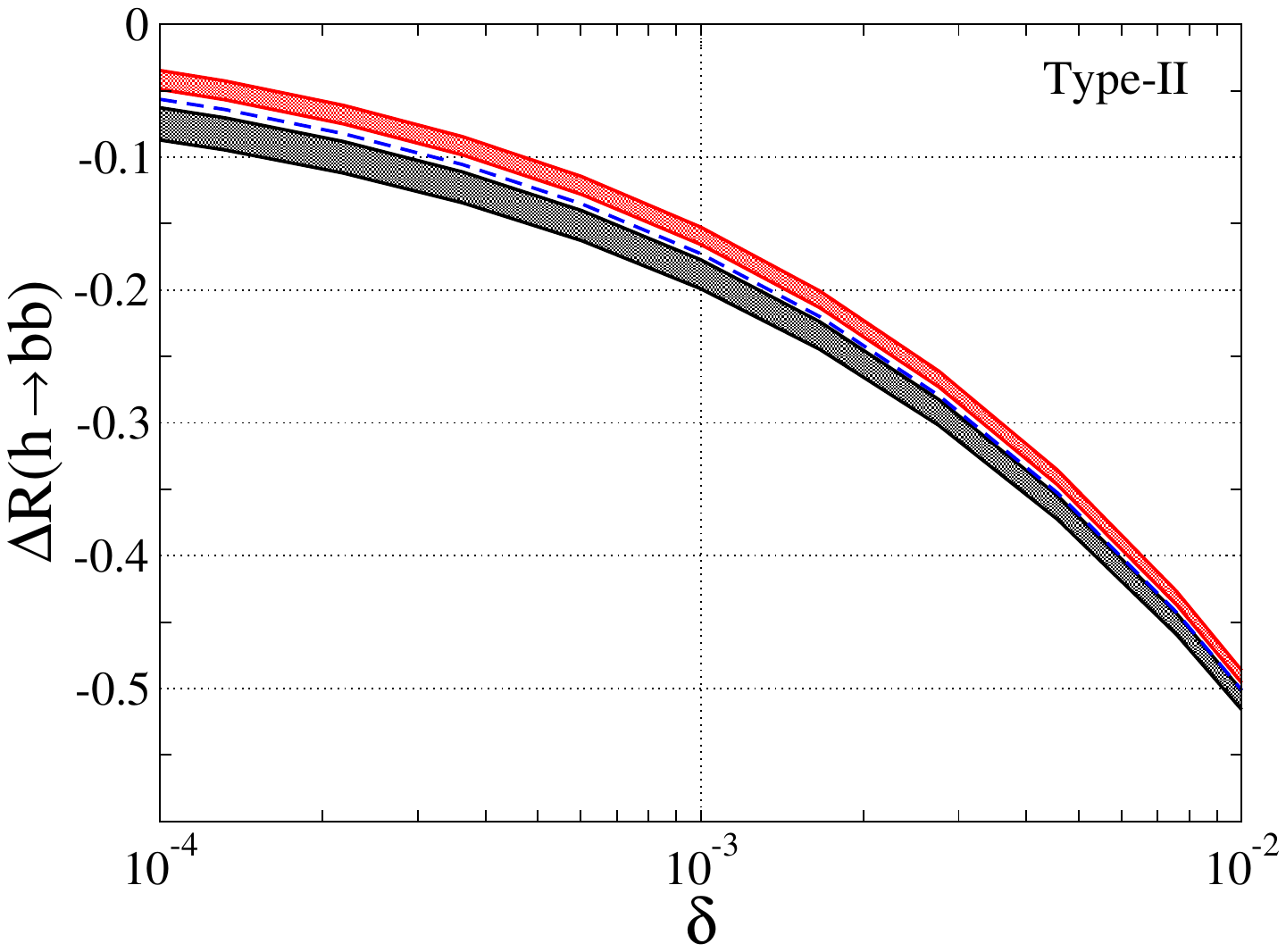}
\includegraphics[width=75mm]{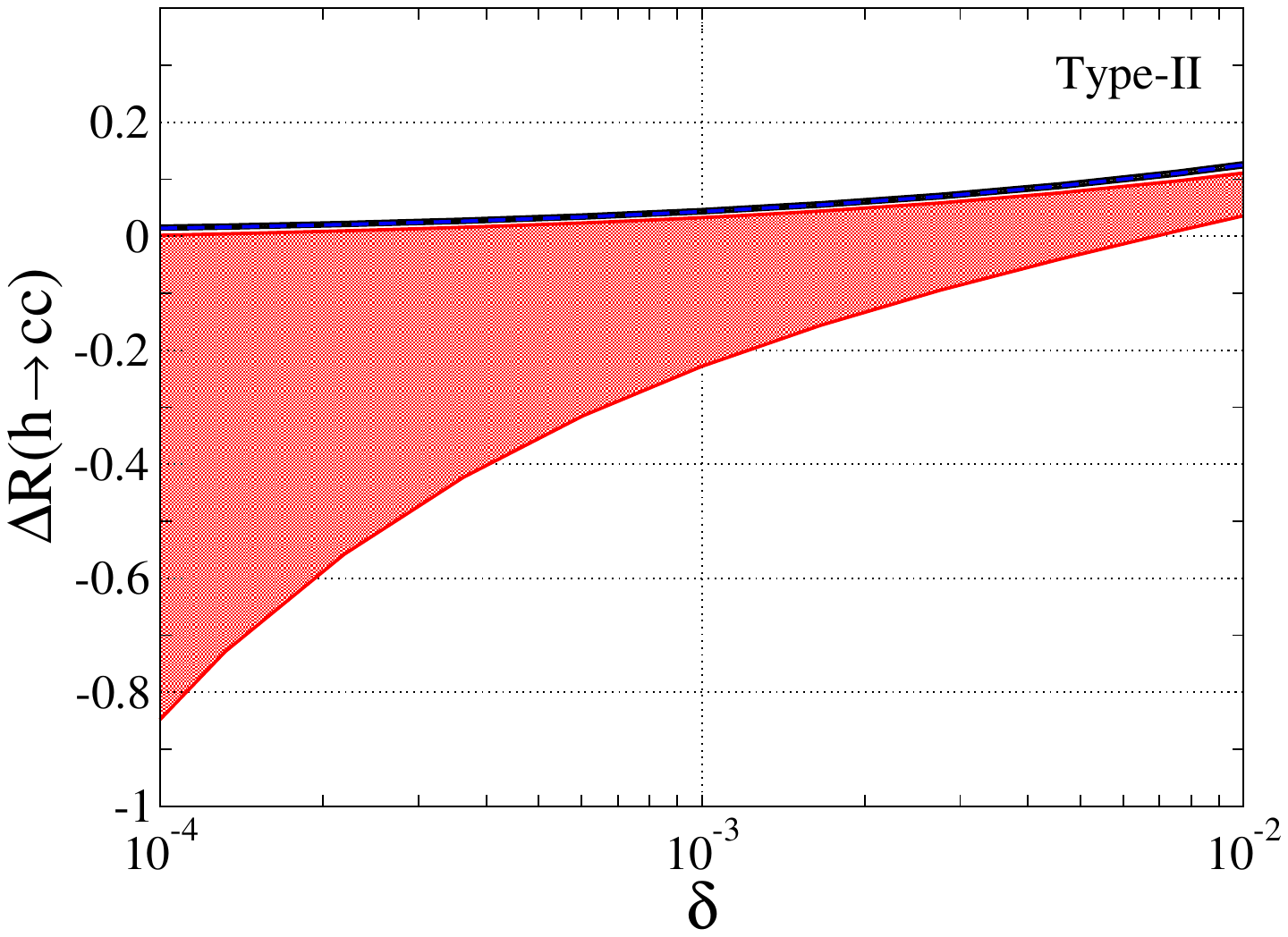}\\
\vspace{-3mm}
\includegraphics[width=75mm]{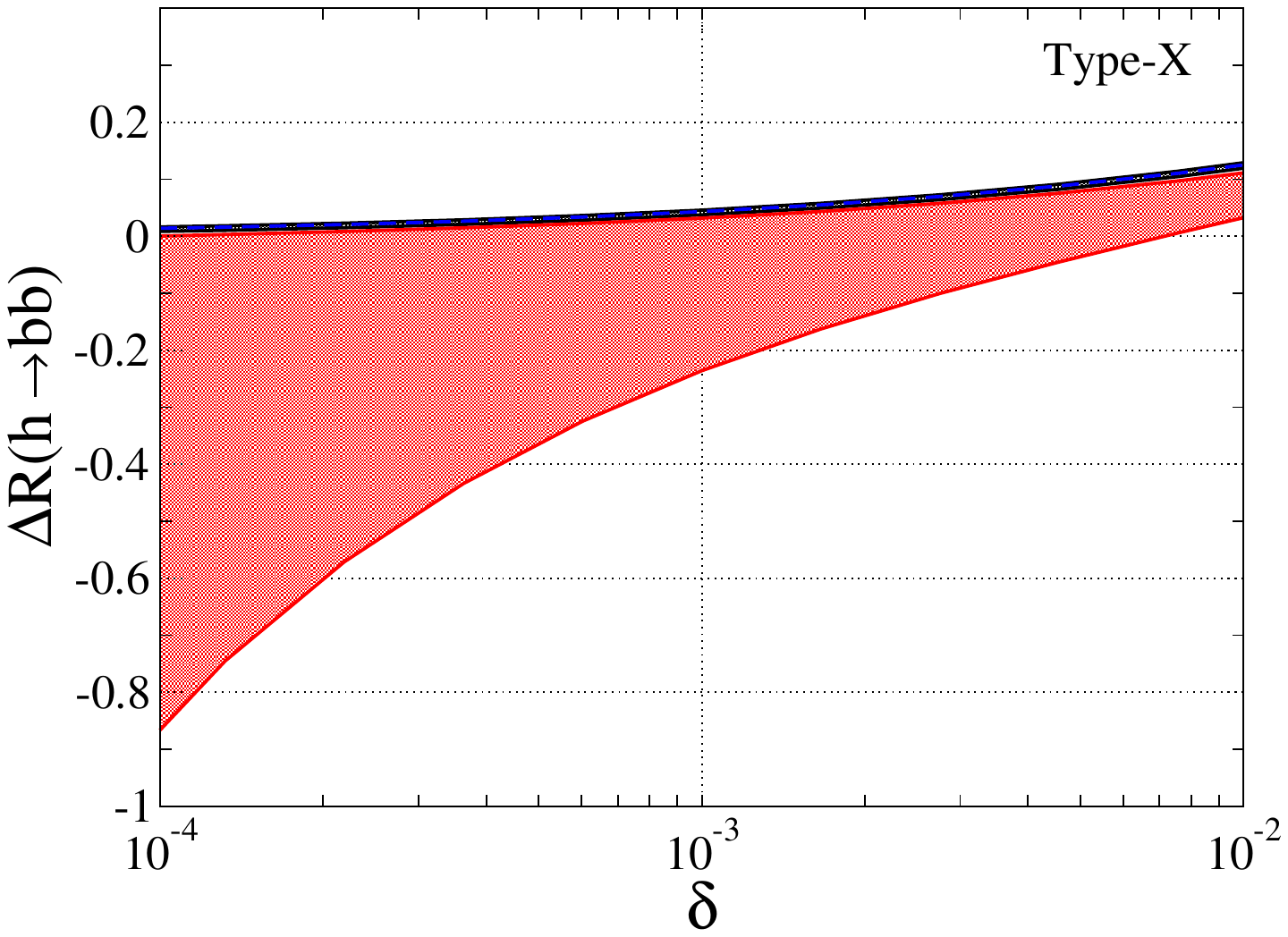}
\includegraphics[width=75mm]{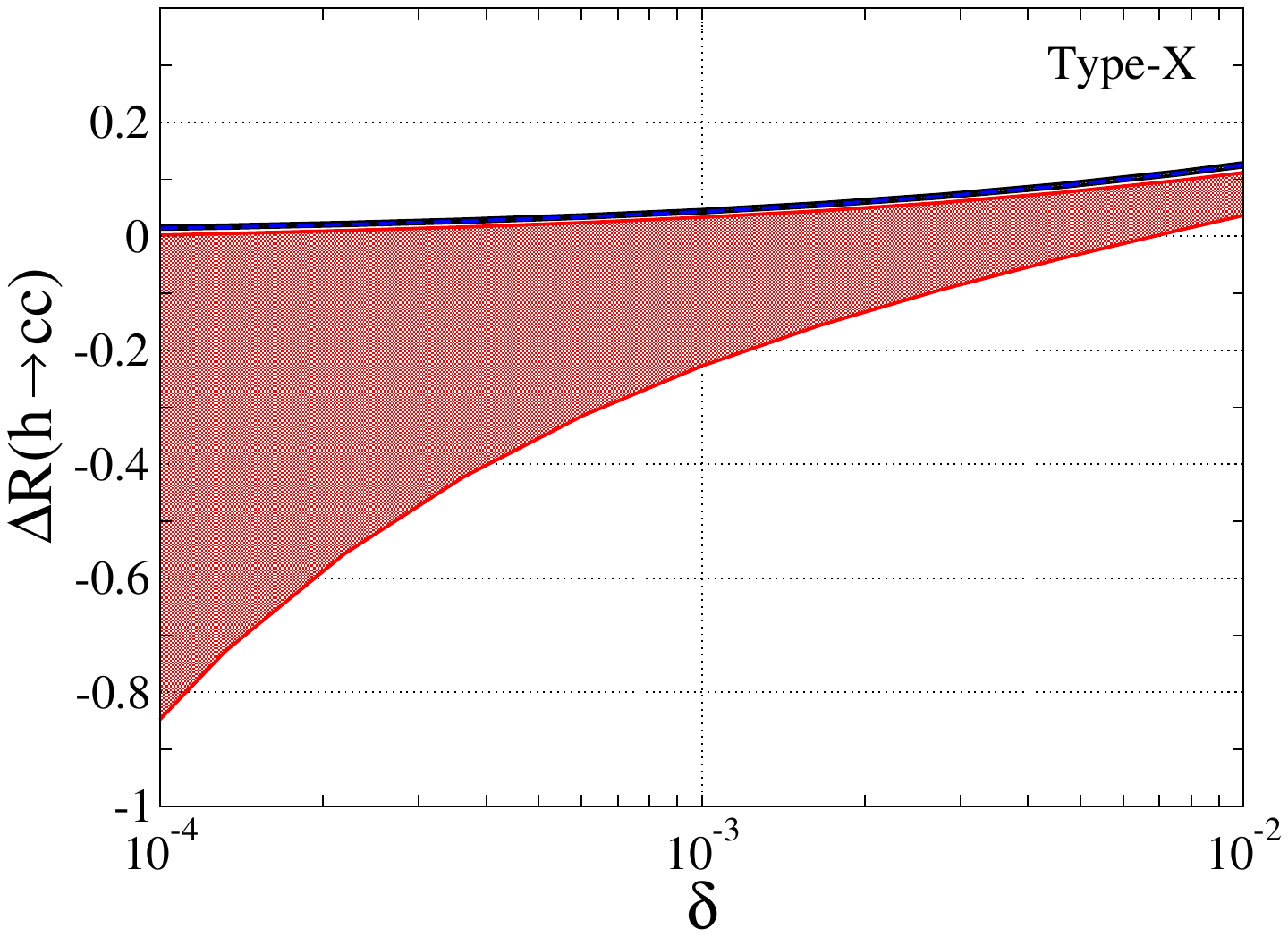}\\
\vspace{-3mm}
\includegraphics[width=75mm]{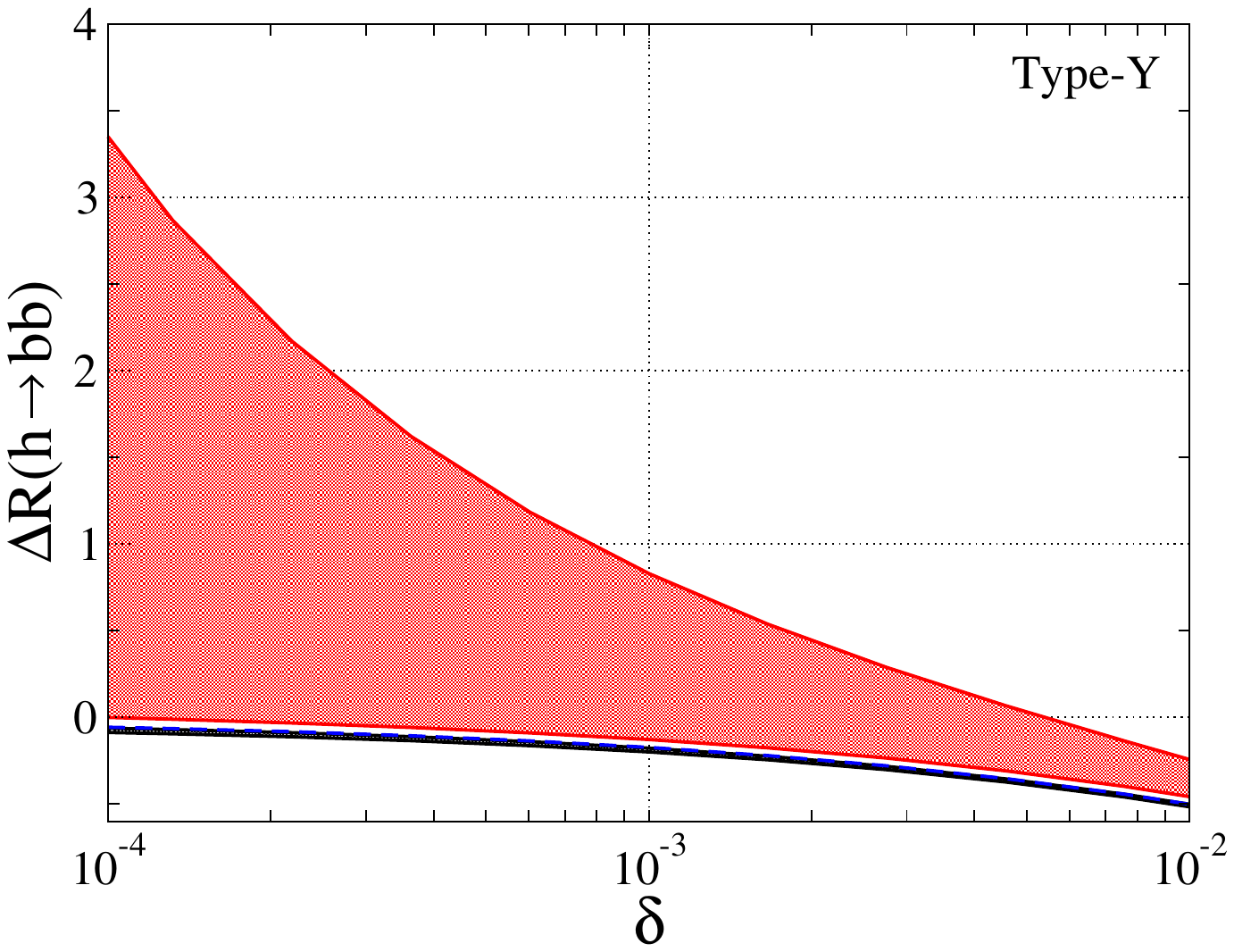}
\includegraphics[width=75mm]{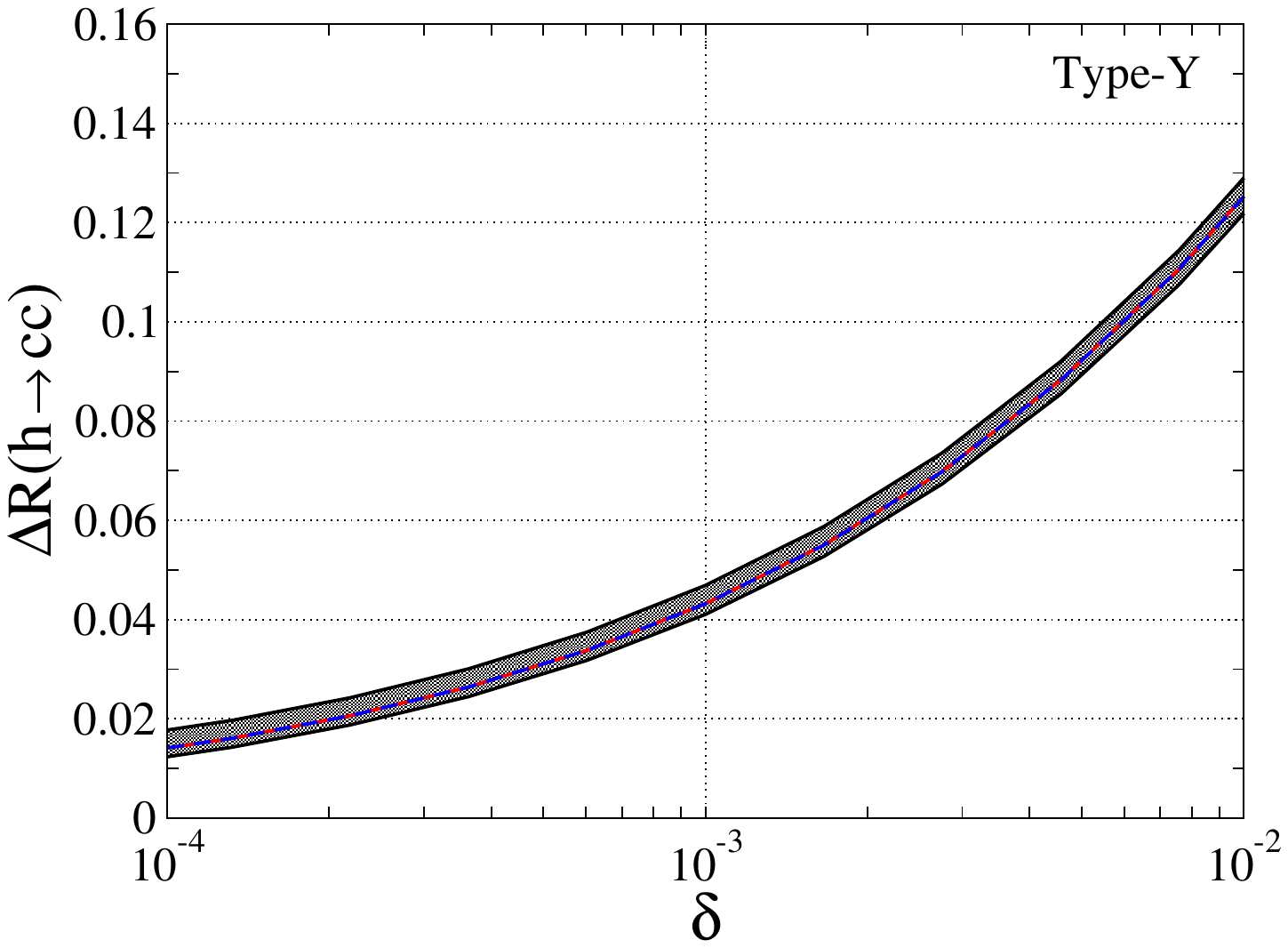}
\end{center}
\vspace{-0.5cm}
\caption{Results of $\Delta R(h \to XY)$ with $XY = b\bar{b}$ (left panels) and $c\bar{c}$ (right panels) as a function of $\delta \equiv 1 - \sin(\beta-\alpha)$ in the Type-II (upper panels), Type-X (middle panles) and Type-Y (bottom panels) 2HDMs. 
In all the panels, we take $m_{H^\pm} = m_H = m_A = 300$ GeV, $\tan\beta = 2$, and scan $M^2$.
The regions black and red shaded respectively show the prediction by using the KOSY scheme and the new scheme. The blue dashed curve shows the prediction at LO. }
\label{fig2}
\end{figure}

In Fig.~\ref{fig2}, we show the results of $\Delta R(h \to q\bar{q})$
with $q = b,c$ in the Type-II, Type-X and Type-Y 2HDMs. 
For the $h \to \tau^+\tau^-$ decay, the result is exactly the same as the LO prediction as we see in Fig.~\ref{fig1}, so we do not show it again. 
Differently from the Type-I 2HDM, the values of $\Delta R$ 
using the new scheme can be largely different from the corresponding value at LO, especially for smaller $\delta$, e.g., $\Delta R(h \to c\bar{c})$
in the Type-II 2HDM.
Such a large difference can be understood by looking at Eq.~(\ref{eq:diff}), which corresponds to the case of $\zeta_f \neq \zeta_\tau$, e.g., $\zeta_\tau\, (=\zeta_e) = -\tan\beta$ and $\zeta_c\, (=\zeta_u) = \cot\beta$ in the Type-II 2HDM. 
We note that if we take $M^2 \simeq m_\Phi^2$ with $m_\Phi^{}$ being the mass of the additional Higgs bosons, then the value of $\Delta R$ takes closer values to those given at LO.  
We note that a larger deviation, e.g., $|\Delta R(h \to f\bar{f})|\sim {\cal O}(10)\%$, can already be excluded by the current LHC data, but we here show such a larger deviation in order to demonstrate the behavior of predictions given in our new scheme.

\begin{figure}[!t]
\begin{center}
\includegraphics[width=100mm]{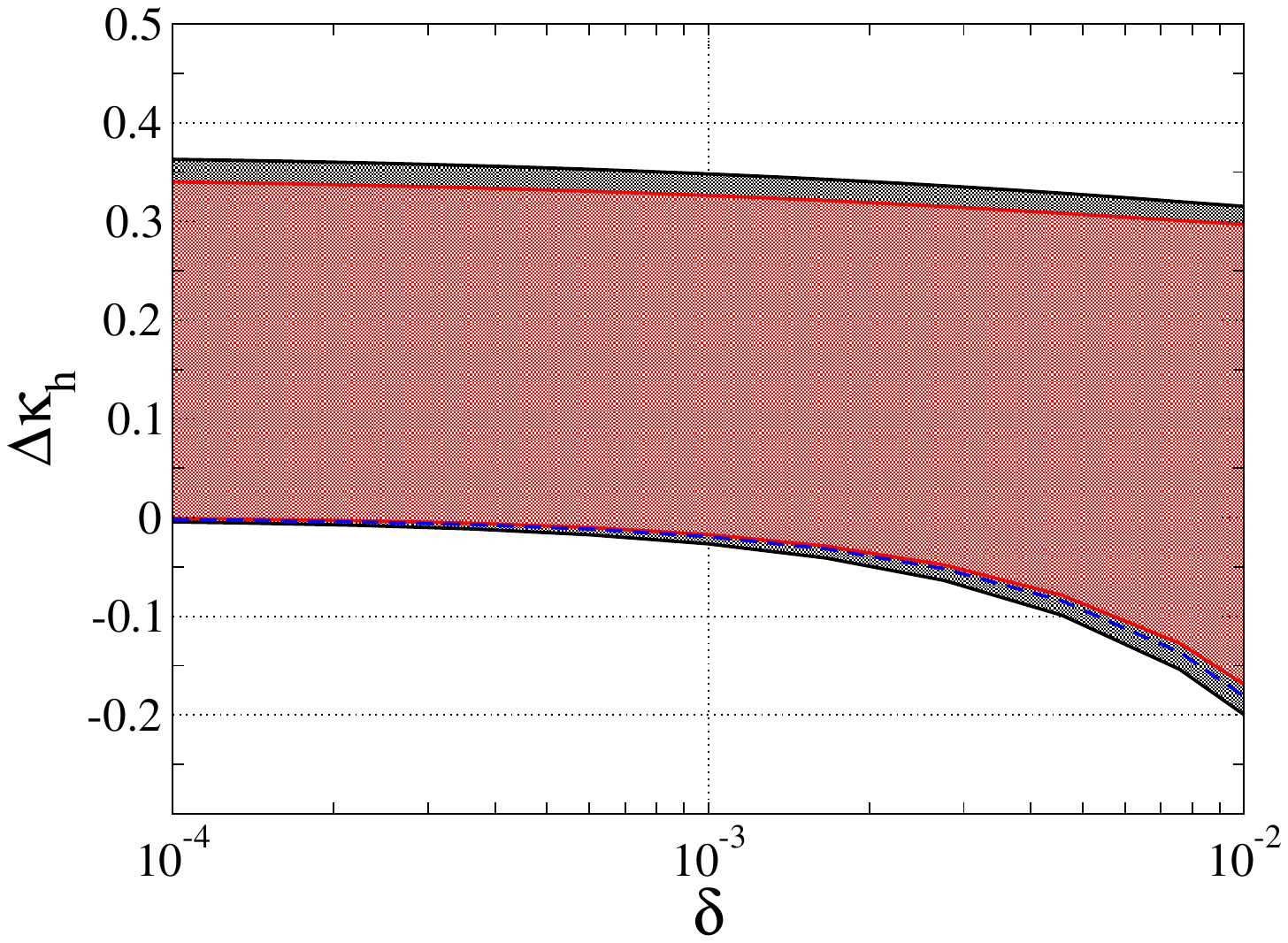}
\end{center}
\caption{Deviation in the $hhh$ coupling $\Delta \kappa_h$ defined in Eq.~(\ref{eq:delkh}) as a function of $\delta$ in the Type-I 2HDM with $m_{H^\pm} = m_A^{} = m_H^{} = 300$ GeV, $\cos(\beta-\alpha) > 0$, $\tan\beta = 2$ and scan $M^2$. We take the renormalization scale $\mu = 300$ GeV.  }
\label{fig3}
\end{figure}

In Fig.~\ref{fig3}, we show the deviation in the $hhh$ coupling from the SM prediction at one-loop level. 
We here define 
\begin{align}
\Delta \kappa_h \equiv \frac{\hat{\Gamma}_{hhh}^{\rm 2HDM}(m_h^2,m_h^2,q^2)}{\hat{\Gamma}_{hhh}^{\rm SM}(m_h^2,m_h^2,q^2)} - 1. \label{eq:delkh}
\end{align}
For the numerical evaluation, we take $q^2 = 260^2$ GeV$^2$. 
As we discussed in Sec.~\ref{sec:reno}, the renormalized $hhh$ vertex is given by using the $\overline{\text{MS}}$ sheme, so that the renormalization scale dependence $\mu$ appears. We here take $\mu= 300$ GeV which is the same as the masses of the additional Higgs bosons. 
We see the large deviation, up to around 35\% from the SM prediction in both the KOSY and new schemes. Such a large deviation has been known as the result of the nondecoupling loop effect of the additional Higgs bosons~\cite{Kanemura:2002vm,Kanemura:2004mg}. \footnote{In Refs.~\cite{Braathen:2019zoh,Braathen:2019pxr}, two-loop calculations of the $hhh$ coupling have been performed, and it further pushes up the deviation in the $hhh$ coupling from the SM prediction as compared with the one-loop result. }
It is clear that the nondecoupling effect does not vanish in the new scheme, but slightly smaller results are obtained compared with the KOSY scheme. 

\begin{figure}[!t]
\begin{center}
\includegraphics[width=75mm]{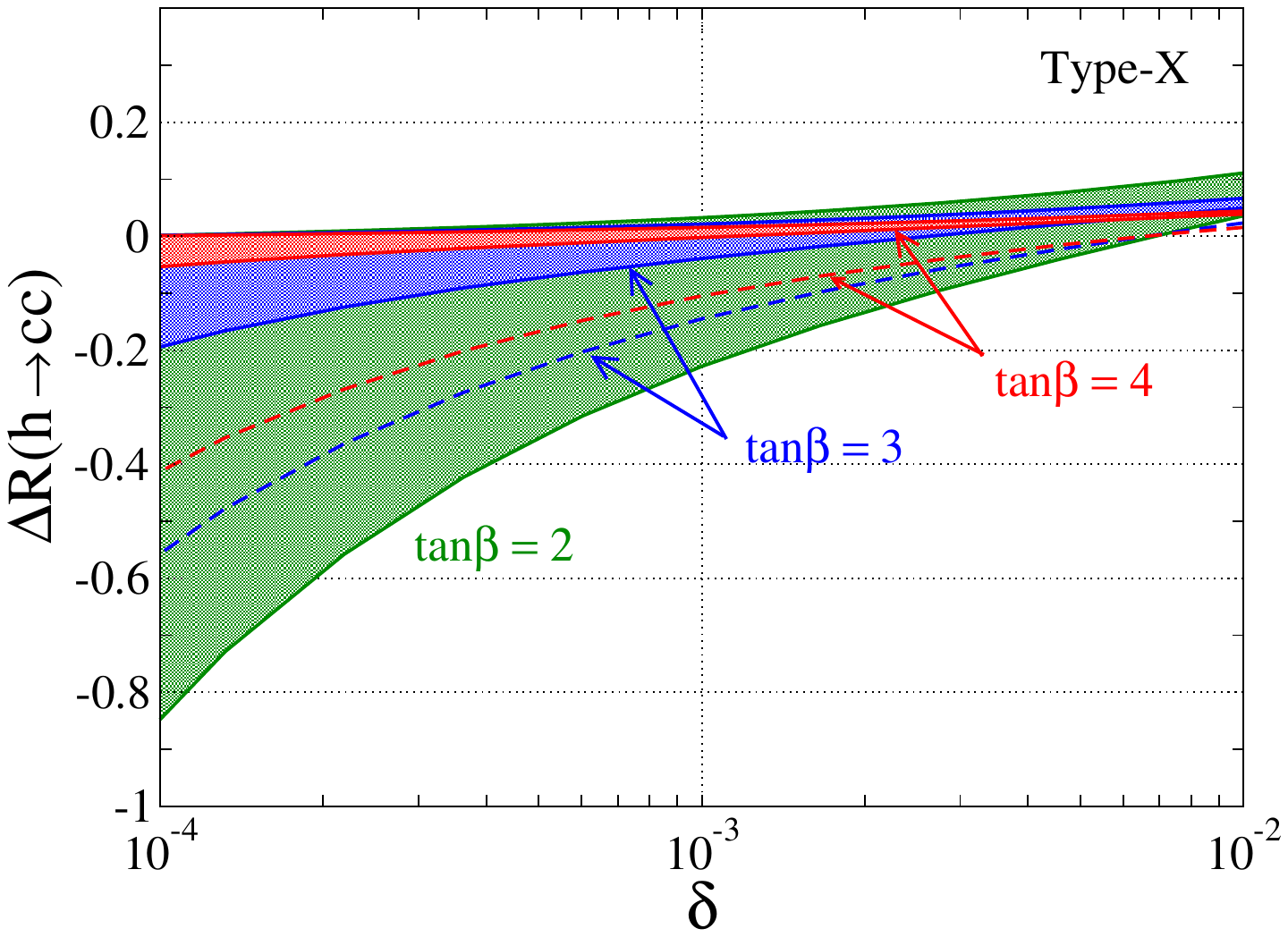}
\includegraphics[width=75mm]{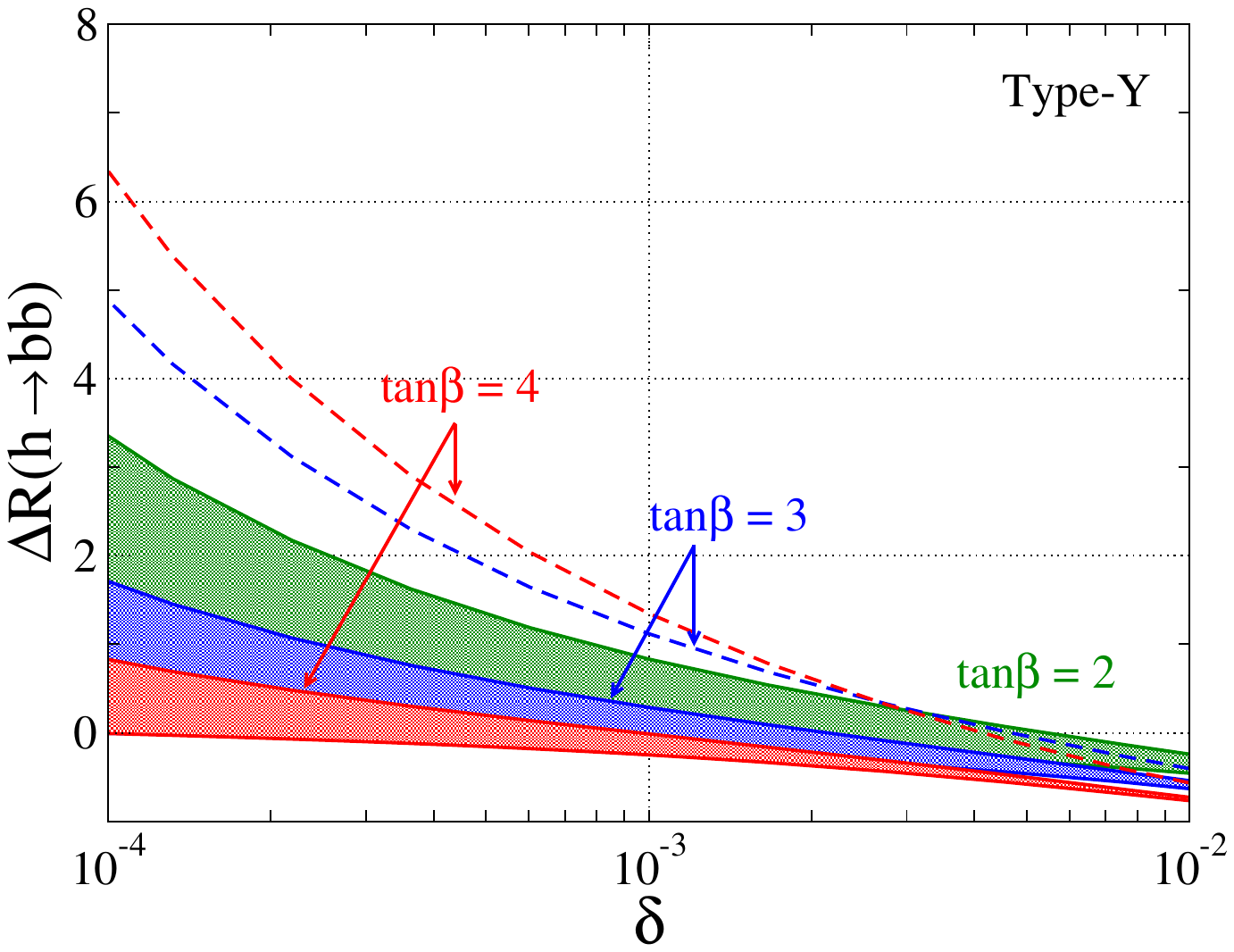}
\end{center}
\caption{
$\Delta R(h \to c\bar{c})$ in the Type-X 2HDM (left) and $\Delta R(h \to b\bar{b})$ in the Type-Y 2HDM (right) as a function of $\delta \equiv 1 - \sin(\beta-\alpha)$. 
In all the panels, we apply the new renormalization scheme, and take $m_{H^\pm} = m_H = m_A = 300$ GeV, $\cos(\beta-\alpha) > 0$ and scan $M^2$. 
The regions shaded with green, blue and red show the case with $\tan\beta = 2$, 3 and 4, respectively. 
For $\tan\beta = 3$ and $4$, we show the prediction with $M^2 = 0$ denoted as the blue dashed and red dashed curve, which are excluded by the bound from the perturbative unitarity. For $\tan\beta = 2$, the case with $M^2 = 0$ is allowed.   }
\label{fig4}
\end{figure}

Let us briefly discuss the case with the other values of $\tan\beta$ in the new scheme. 
For the case with $\zeta_f = \zeta_\tau$, the deviation in the decay rate of $h \to f\bar{f}$ is roughly given by its LO results. 
For instance, $|\Delta R(h \to f\bar{f})|$ tends to be smaller for larger $\tan\beta$ in the Type-I 2HDM, because the scaling factor $\kappa_f^h$ gets closer to unity for fixed $\delta$. 
For $\zeta_f \neq \zeta_\tau$, we show how the results are changed depending on the value of $\tan\beta$ in Fig~\ref{fig4}. 
We here particularly demonstrate $\Delta R (h \to c\bar{c})$ in the Type-X 2HDM
and $\Delta R (h \to b\bar{b})$ in the Type-Y 2HDM, because the other cases, e.g., 
$\Delta R(h \to c\bar{c})$ in the Type-II 2HDM and $\Delta R(h \to b\bar{b})$ in the Type-X 2HDM show the similar behavior to $\Delta R (h \to c\bar{c})$ in the Type-X 2HDM. 
We see that $|\Delta R (h \to c\bar{c})|$ in the Type-X 2HDM becomes smaller 
for larger $\tan\beta$ due to the two reasons. 
First, the magnitude of the coefficient of $t_{\beta-\alpha} \delta Z_h $ given in Eq.~(\ref{eq:diff}) takes smaller values for larger $\tan\beta$. 
Second, smaller values of $M^2$ are excluded by the bound from the perturbative unitarity, and then the amount of the nondecoupling effect is suppressed.      
In fact, by comparing the dashed curves (results with $M^2 = 0$) and the lower edge of the blue- and red-shaded regions, we see that the prediction is shrunk due to the unitarity bound.  
On the other hand, $|\Delta R (h \to b\bar{b})|$ in the Type-Y 2HDM become larger for larger $\tan\beta$ if we ignore the unitarity bound (dashed curves). 
However, if we take into account the bound, maximally allowed values of $|\Delta R (h \to b\bar{b})|$ are smaller than that for $\tan\beta = 2$. 

Before closing this section, let us also comment on the case with $c_{\beta-\alpha} < 0$, while we took $c_{\beta-\alpha} > 0$ in the above. 
In this case, $\Delta R$ shows similar behavior to that with $c_{\beta-\alpha}>0$ as discussed above, but of course the scaling factors $\kappa_f^h$ take different values depending on the sign of $c_{\beta-\alpha}$, so that the prediction of $\Delta R(h \to f\bar{f})$ at NLO is changed according to the change of $\kappa_f^h$.  
For instance, the sign of $\kappa_f^h - 1$ tends to be flipped between two cases with $c_{\beta-\alpha} > 0$ and $c_{\beta-\alpha} < 0$, so that the sign of $\Delta R(h \to f\bar{f})$ also tends to be flipped between two cases. 


\section{Discussion and Conclusion}

Let us give a few remarks on the discussion given in the present Letter.  
First, in the new scheme, we cannot take the exact alignment, i.e., $\sin(\beta - \alpha) = 1$, because the dependence of $\delta \alpha$ and $\delta \beta$ vanishes in the counterterms of the $hZZ$ and $h\tau^+\tau^-$ vertices. We can still take the alignment limit $\sin(\beta -\alpha) \to 1$, where these two mixing counterterms can be defined. 
This statement can generally be true in extended Higgs models with a mixing of Higgs bosons such as the Higgs singlet model (a model with an additional isospin singlet scalar field) \footnote{In the Higgs singlet model, only one mixing parameter appears in the Higgs boson couplings~see e.g.,~\cite{Kanemura:2016lkz}, so that we can impose one of two renormalization conditions to determine the corresponding counterterm.}, multi doublet models (with more than two Higgs doublets) and models with higher isospin representations, e.g., with triplets. 
In this class of models, we can apply the new scheme, by which some of the decay rates of $h$ are taken to be the input values as we have shown in the 2HDM as an example.  

%

Second, we can realize the exact alignment in models with an unbroken symmetry in which new scalar fields are charged under the symmetry and they do not mix with the Higgs doublet field giving masses of SM particles. 
The inert doublet model is a simple example of such a model, where
the second doublet field is assigned to be odd under an unbroken $Z_2$ symmetry. 
In this class of models, the renormalization of the Higgs sector can be done in a similar way to that of the SM. Namely, the renormalized vertices for $h$ do not contain additional counterterms, so that the same set of the renormalization conditions as in the SM is enough to be imposed for the renormalization of these vertices.  
Therefore, radiative corrections to the decay rates of $h$ cannot be absorbed by the renormalization, and information on inert particles can be extracted from the loop effects.

We have discussed a new renormalization scheme in the 2HDMs with a softly-broken $Z_2$ symmetry and CP-conservation in the Higgs sector.
In this scheme, we have applied the decay rates of the discovered Higgs boson $h$
into $ZZ^* (\to Z\ell^+\ell^-)$ and $\tau^+\tau^-$ at NLO 
to the renormalization condition such that their values take the corresponding prediction at NLO in the SM times scaling factor squared at tree level. 
This means that the scaling factors still work to describe the alignmentness, i.e., how the predictions of $h$ are close to the SM predictions, at loop levels. 
We have derived the expressions for two counterterms $\delta \alpha$ and $\delta \beta$ by using the above renormalization conditions, and then formulated the 
electroweak corrections to the decay rates of $h$ at NLO.

We have compared the deviations in the decay rates of $h$ at NLO from the corresponding SM predictions in the new scheme with those calculated in the previous scheme, the so-called KOSY scheme.  
We have confirmed that the deviation in the decay rates of $h \to ZZ^*$ and $h \to \tau^+\tau^-$ at NLO well agree with the corresponding LO results as anticipated. 
For $h \to WW^*$, the prediction at NLO differs from the LO results, but its amount is of order $10^{-3}$-$10^{-4}$. 
For $h \to f\bar{f}$ ($f \neq \tau$), the situation strongly depends on the type of Yukawa interactions and the type of fermions.  
For instance, if we consider the Type-I 2HDM, all the decay rates of $h \to f\bar{f}$ take closer values to the LO results than those given in the KOSY scheme. 
On the other hand, for the other types e.g., Type-II 2HDM, 
the decay rate of $h \to c\bar{c}$ can be largely different from the LO results in the new scheme. This comes from the fact that the mixing factor, denoted as $\zeta_\tau (=\zeta_e)$, is different from $\zeta_f$ $(f \neq \tau)$, which gives rise to the dependence of $\tan(\beta-\alpha)$ in the electroweak corrections to the decay rate. 
Finally, we have computed the one-loop corrections to the $hhh$ coupling in both the new scheme and the KOSY scheme, and have found that both the schemes give a sizable deviation from the SM prediction due to the nondecoupling loop effect of the additional Higgs bosons. As an example, the amount of the deviation can be 30\%-40\% when the masses of the additional Higgs bosons are taken to be 300 GeV.   

In conclusion, by using the new scheme proposed in this Letter, we can input the precisely measured values of the decay rates of $h \to ZZ^*$ and $h \to \tau^+\tau^-$ at future Higgs factories, and then we check if the model is consistent or not by comparing 
the predictions of the other decay rates with the corresponding precise measurements.   

\begin{acknowledgments}
The authors would like to thank Kodai Sakurai and Masashi Aiko for useful discussions. 
This work is supported in part by the Grant-in-Aid on Innovative Areas, the Ministry of Education, Culture, 
Sports, Science and Technology, No.~20H00160 and No.~23K17691 [S.K.],
Early-Career Scientists, No.~20K14474 [M.K.].
\end{acknowledgments}

\vspace*{-4mm}

\bibliography{references}

\end{document}